\documentclass[twoside]{article}

\usepackage[accepted]{aistats2024}
\usepackage[round]{natbib}

\usepackage{amsgen,amsmath,amstext,amsbsy,amsopn,amssymb,mathabx,amsthm,bm,graphicx,breqn,dsfont}
\usepackage{xr}
\usepackage[utf8]{inputenc} 
\usepackage[T1]{fontenc}    
\usepackage{hyperref}       
\usepackage{url}            
\usepackage{booktabs}       
\usepackage{amsfonts}       
\usepackage{microtype}      
\usepackage{xcolor}         

\usepackage{float}
\usepackage{multirow,bigstrut,makecell}
\usepackage[ruled,vlined]{algorithm2e}
\usepackage{wrapfig}
\usepackage{caption}
\usepackage{subcaption}

\newtheorem{thm}{Theorem}[section]
\newtheorem{lem}[thm]{Lemma}

\theoremstyle{definition}

\newtheorem{asp}{Assumption}

\theoremstyle{remark}

\newcommand{\bbf}{\bm{f}}

\newcommand{\bx}{\bm{x}}

\newcommand{\bX}{\bm{X}}

\newcommand{\bZ}{\bm{Z}}

\newcommand{\Lb}{\mathbf{L}}

\newcommand{\bbE}{\mathbb{E}}

\newcommand{\bbP}{\mathbb{P}}

\newcommand{\bbR}{\mathbb{R}}

\newcommand{\cA}{\mathcal{A}}

\newcommand{\cD}{\mathcal{D}}

\newcommand{\cH}{\mathcal{H}}

\newcommand{\cR}{\mathcal{R}}
\newcommand{\cS}{\mathcal{S}}

\newcommand{\cV}{\mathcal{V}}

\newcommand{\cX}{\mathcal{X}}

\newcommand{\sgn}[1]{\operatorname{sign} \{#1\}}
\DeclareMathOperator*{\argmin}{arg\,min}

\newcommand{\bbone}{{\mathds{1}}}  

\newcommand{\norm}[1]{\left\lVert#1\right\rVert}

\newcommand{\abs}[1]{\left\lvert#1\right\rvert}
\newcommand{\absn}[1]{\lvert#1\rvert}
\newcommand{\prth}[1]{\left(#1\right)}
\newcommand{\brck}[1]{\left[#1\right]}
\newcommand{\brce}[1]{\left\{#1\right\}}

\begin{document}

%

%

\twocolumn[

\aistatstitle{Fusing Individualized Treatment Rules Using Secondary Outcomes}

\aistatsauthor{ Daiqi Gao \And Yuanjia Wang \And Donglin Zeng }

\aistatsaddress{ Harvard University \And  Columbia University \And University of Michigan } ]

\begin{abstract}
    An individualized treatment rule (ITR) is a decision rule that recommends treatments for patients based on their individual feature variables. In many practices, the ideal ITR for the primary outcome is also expected to cause minimal harm to other secondary outcomes. Therefore, our objective is to learn an ITR that not only maximizes the value function for the primary outcome, but also approximates the optimal rule for the secondary outcomes as closely as possible. To achieve this goal, we introduce a fusion penalty to encourage the ITRs based on different outcomes to yield similar recommendations. Two algorithms are proposed to estimate the ITR using surrogate loss functions. We prove that the agreement rate between the estimated ITR of the primary outcome and the optimal ITRs of the secondary outcomes converges to the true agreement rate faster than if the secondary outcomes are not taken into consideration. Furthermore, we derive the non-asymptotic properties of the value function and misclassification rate for the proposed method. Finally, simulation studies and a real data example are used to demonstrate the finite-sample performance of the proposed method.
\end{abstract}

\section{INTRODUCTION}

An individualized treatment rule (ITR) is a decision rule that recommends treatments to patients based on their pre-treatment covariates, such as demographics, medical history, and health status.
Many methods have been proposed to estimate the optimal ITR using data from a randomized controlled trial or an observational study. 
The goal is to find the ITR that would yield the maximal primary outcome if patients follow the treatment recommendations.
Regression-based approaches start with estimating Q-functions, which are the expected outcomes associated with each treatment option, and the optimal treatment is determined by maximizing the Q-functions. 
Exemplary methods include Q-learning \citep{qian2011performance,ma2022learning} and A-learning \citep{murphy2003optimal,shi2018high}. 
Alternative methods directly search for the ITR that maximizes the mean reward, also called the value function, using either an inverse probability weighted estimator \citep{zhao2012estimating,chen2016personalized,zhang2020multicategory,gao2022non,ma2023learning} or a doubly robust estimator \citep{zhang2012robust,zhou2017residual,liu2018augmented,zhao2019efficient}. 
However, all these methods focus on one primary outcome when estimating the ITRs.

In practical settings, it is important to consider certain secondary outcomes. 
Non-favorable secondary outcomes potentially indicate worsened overall health or an increased risk of treatment non-compliance. 
Therefore, when estimating the optimal ITR for the primary outcome, it is crucial to ensure that the ITR also optimizes the secondary outcomes to the greatest extent. 
For example, for major depressive disorder  \citep[MDD;][]{trivedi2016establishing}, one common outcome to measure depressive symptoms is the Quick Inventory of Depressive Symptomatology (QIDS) score, a rating scale of the patient's emotional state over the past seven days. 
Besides, another important outcome is the Clinical Global Improvement (CGI) scale, which is often used to assess a patient's symptoms, behavior, and the impact on the patient's ability to function. It is an indicator of overall clinical improvement.
Although the primary goal in many studies is to identify the optimal treatment strategy for improving the QIDS score, it is equally important for such a strategy to be effective for the CGI scale.

In this work, we have a twofold objective, which is to estimate the optimal ITR for the primary outcome and to ensure that the resulting treatment decision closely aligns with the optimal treatment for the secondary outcomes. 
Specifically, we aim to fuse the treatment rules for different outcomes to provide a unified ITR that performs optimally for the primary outcome and effectively for the secondary outcomes, even if it may not be optimal for the latter.
To achieve this goal, we introduce a novel fused learning framework to estimate the optimal ITR, namely the fused individualized treatment rule (FITR). 
Under this framework, we maximize the value function for the primary outcome.
At the same time, we incorporate a fusion penalty to encourage agreement between the estimated ITR and the optimal ITRs for the secondary outcomes. The latter are assumed to be obtained apriori using external data or the data from the same study. 
The fusion penalty is designed as a weighted sum of the disagreement rates between the treatment rules, with weights determined based on the similarity of treatment effect on these outcomes. 
Therefore, this penalty aims to encourage consistency between the ITRs for the different outcomes.

\paragraph{Related Work.}
Several approaches have been proposed to learn ITRs that can handle multiple outcomes simultaneously.
Some of these approaches \citep{wang2018learning,laber2018identifying,fang2022fairness} estimated the optimal ITRs for the primary outcome while restricting the secondary outcome to be no less (or no larger if the outcome is a risk outcome) than a threshold. 
However, these approaches require the threshold to be pre-specified and only ensure that the average value of the secondary outcome is controlled, but not for any specific individuals. 
Another class of methods 
\citep{thall2002selecting,luckett2021estimation} constructed expert-derived or data-driven patient-specific composite outcomes for learning the ITRs. 
However, their ITRs may not be optimal for the primary outcome of interest.
In addition, \citet{laber2014set} and \citet{lizotte2016multi} proposed constructing a set-valued regime for competing outcomes to generate non-inferior outcome vectors but does not recommend a single treatment, limiting their use for practice.
Our goal is to learn an optimal ITR for the primary outcome while ensuring closeness to the optimal secondary outcomes, which is distinct from all existing approaches for learning multi-outcome ITRs. 

Our problem is also fundamentally different from the literature that combines multiple studies in analysis.
Meta-analysis \citep{haidich2010meta,lin2010relative,claggett2014meta,liu2015multivariate} combines the estimators across multiple studies efficiently.
Integrative data analysis \citep{curran2009integrative,brown2018two} considers multiple data sources or summary statistics using shared parameter models.
Transfer learning \citep{li2021transfer,cai2021transfer,tian2022transfer} uses existing knowledge from one task to assist in learning a new task. 
However, these existing methods often require individual-level data or assume specific models or distributions for each study to achieve integration, so are not applicable to combining different treatment rules. 


\paragraph{Main Contributions.}
Our paper introduces several significant contributions. 
(1) We propose a fusion penalty to encourage the primary outcome ITR to closely agree with the optimal secondary outcome ITRs.
The proposed method relies on the secondary outcome only through its treatment rules. 
We do not require individual-level data for the secondary outcome when such rules are already available. 
In addition, there is no restriction on the form of the decision function since the fusion penalty is directly applied to the disagreement rate. 
The proposed method is not affected by the magnitude of the decision function or the function space in which the secondary outcome ITRs are estimated.
(2) To efficiently solve for the FITR, we propose two algorithms that use surrogate losses to substitute the noncontinuous, nonconvex value function and fusion penalty in the objective function.
(3) Theoretically, we prove that the agreement rate between the estimated FITR learned with the surrogate loss will converge to the true agreement rate, and the convergence rate is faster than the approach without considering the secondary outcome. 
We also obtain the convergence rate for the value function and misclassification rate of FITR. 
(4) In the numeric experiments, we show that the agreement rate indeed converges faster to the true rate. 
Moreover, when the true optimal ITRs of different outcomes are closely aligned, the learned FITR actually yields higher value function and accuracy for the primary outcome compared to traditional methods. 
By leveraging the secondary outcomes as useful side information, the proposed method has the capability to improve the treatment decisions for multiple outcomes simultaneously.


\section{METHODOLOGY} \label{sec:methodology}

The vector $\bX \in \cX \subset \bbR^d$ represents pre-treatment covariates for tailoring treatment decisions, where $d$ is the dimension of $\bX$. 
The treatment $A \in \cA = \brce{1, -1}$ is assumed to be binary.
The primary outcome is denoted by $R_1$. Without loss of generality, we assume that a higher value of this outcome indicates a better health condition.
Denote $R_1(a)$ as the potential outcome under the treatment $a$.
For the primary outcome $R_1$, an ITR $\cD_1: \cX \to \cA$ maps a patient's covariates to a treatment suggestion. 
It can also be expressed as $\cD_1(\bX)=\sgn{f_1(\bX)}$ for some decision function $f_1: \cX \to \bbR$.
Define $\cV_1 (f_1) := \bbE [R_1 \{ \cD_1(\bX) \}]$ as the value function associated with $f_1$, which is the average potential outcome when the treatments follow the ITR $\cD_1 = \sgn{f_1}$ for all $\bX$. 
Our goal is to estimate the optimal ITR that maximizes this value function.
The optimal ITR $\cD_1^*$ is given as $\sgn{f_1^*}$, where 
$f_1^*(\bx)=\bbE [R_1(1) | \bX = \bx] - \bbE [R_1(-1) | \bX = \bx]$.

We make standard assumptions in causal inference, including ignorability, consistency, and positivity (see Section~\ref{suppsec:sub.theory}), so that the optimal ITR can be estimated from a sample of $n$ i.i.d data of $(\bX, A, R_1)$. 
Under these conditions, according to \citet{qian2011performance}, we have $\cV_1 (f_1)=\bbE[R_1 \bbone \{Af_1(\bX)>0 \}/\pi(A;\bX)]$,
where $\pi(A;\bX)$ is the probability of taking treatment $A$ given covariates $\bX$ in the data. 
Thus, the optimal ITR can be estimated by solving
\begin{equation*}
\label{equ:f.optimal}
    \min_{f_1 \in \cH} \frac{1}{n} \sum_{i=1}^n \frac{R_{i1} \bbone \{A_i f_1 (\bX_i) < 0 \}}{\pi (A_i; \bX_i)} + \lambda_{1n} \norm{f_1}^2,
\end{equation*}
where $\cH$ is some function class, $\norm{f_1}$ is the semi-norm for $f_1$ in its function space, and $\lambda_{1n}$ is a tuning parameter that depends on the sample size $n$.  

\paragraph{Separate Learning (SepL)}
Since minimizing a loss function with 0-1 loss is computationally challenging, it can be substituted by some convex surrogate loss \citep{bartlett2006convexity}, denoted by $\phi(x)$. 
For example, 
\citet{zhao2012estimating} proposed using the hinge loss where $\phi(x)=\max(1-x,0)$. 
In our implementation, we propose using the logistic loss, $\phi(t) = \log (1 + e^{-t})$, due to its differentiability and computational stability.
To further reduce the variability of the value function estimator, we replace $R_{i1}$ by $R_{i1} - \bbE(R_{i1} | \bX_i)$, which does not change the optimal ITR.
The conditional expectation $\bbE(R_{i1} | \bX_i)$ can be estimated by fitting $R_{i1}$ against $\bX_i$ with any regression methods, e.g. simple linear regression.
Moreover, we simultaneously flip the sign of $R_{i1} - \bbE(R_{i1} | \bX_i)$ and  $A$ when $R_{i1} - \bbE(R_{i1} | \bX_i)$ is negative (c.f., \citealp{liu2018augmented}).
This allows us to formulate the problem as a convex optimization problem
\begin{equation*}
\label{equ:fn.hat}
\begin{split}
    \widetilde{f}_{1n} 
    & = \argmin_{f_1 \in \cH} \frac{1}{n} \sum_{i=1}^n \frac{\abs{R_{i1} - \bbE(R_{i1} | \bX_i)}}{\pi(A_i; \bX_i)} \cdot \\
    & \phi [A_i \sgn{R_{i1} - \bbE(R_{i1} | \bX_i)} f_1 (\bX_i)] 
    + \lambda_{1n} \norm{f_1}^2.
\end{split}
\end{equation*}
We refer to this method as separate learning (\textit{SepL}).

\subsection{Learning Fused ITR Using Optimal Rules for Secondary Outcomes} \label{sec:sub.FITR.secondary}

The secondary outcomes are denoted as $R_2, \dots, R_K$ for $K \ge 2$.
Suppose we have obtained the ITRs $\widetilde{f}_{2}, \dots, \widetilde{f}_{K}$ for the secondary outcomes either from external data or the same study.
Our goal is to estimate the optimal ITR for the primary outcome while encouraging it to be consistent as much as possible with these secondary outcome ITRs. 
To this end, we propose a \textit{fusion penalty} on the disagreement rates between $f_1$ and $\widetilde{f}_{2}, \dots, \widetilde{f}_{K}$.
Specifically, we estimate the \textit{fused individualized treatment rule (FITR)} $f_1$  by minimizing
\begin{equation} \label{equ:fkn.hat}
\begin{split}
    \frac{1}{n} \sum_{i = 1}^n \frac{R_{i1}}{\pi(A_i; \bX_i)} \bbone \{A_i f_1 (\bX_i) < 0\} + \lambda_{1n} \norm{f_1}^2 \\
    + \frac{\mu_{1n}}{n} \sum_{i = 1}^n \sum_{k=2}^K \Omega_{1k} \bbone \{f_1 (\bX_i) \widetilde{f}_k (\bX_i) < 0\},
\end{split}
\end{equation}
where $\Omega_{1k}$ is a nonnegative pre-specified constant that reflects the prior knowledge of the similarity between the primary outcome and each secondary outcome $R_{k}$. 
For example, $\Omega_{1k}$ can be expert-specified or defined as the correlation between $R_1$ and $R_k$.
If $\Omega_{1k}$ is negative, we can simultaneously flip the sign of $\Omega_{1k}$ and $\widetilde{f}_k$ to encourage the consistency between $f_1$ and $-\widetilde{f}_k$.
The tuning parameter $\mu_{1n}$ of the fusion penalty, which reweights $\Omega_{1k}$ for all $k$, is determined data-adaptively using cross-validation. 
It is selected to maximize the estimated value function of $f_1$, thus automatically quantifying the similarity between the outcomes.

The fusion penalty is related to the Laplacian penalty \citep{huang2011sparse}, which is used to learn multiple models simultaneously while encouraging similarity among them.
With $\Lb$ as the Laplacian matrix and $\bbf := \{ f_1, \dots, f_K \}$ as the vector of decision functions for each outcome, the Laplacian penalty can be defined as
        $\frac{1}{n} \sum_{i = 1}^n [ \bbone \{\bbf (\bX_i) > 0\} \cdot \Lb \cdot \bbone \{\bbf (\bX_i) > 0\}^T$ 
        $+ \bbone \{\bbf (\bX_i) < 0\} \cdot \Lb \cdot \bbone \{\bbf (\bX_i) < 0\}^T ]$.
It prompts $f_k (\bX_i)$ and $f_j (\bX_i)$ to have the same sign when $L_{kj} < 0$ for $k, j = 1, \dots, K, k \ne j$.
This is equivalent to (\ref{equ:fkn.hat}) if we use the same data to learn FITRs for all outcomes simultaneously.

\paragraph{FITR-Ramp.}
The optimization problem in (\ref{equ:fkn.hat}) is known to be NP-hard. To tackle this problem, we substitute the 0-1 losses with a smooth loss function for optimization purposes.
The 0-1 loss in the first part of the expression is substituted by a logistic loss as in SepL. 
For the 0-1 loss in the fusion penalty,
we use a ramp loss $\psi_{\kappa} (t) = \min \brce{1, \max \brce{0, 1 - t / \kappa}}$, where $\kappa$ is a tuning parameter, for approximation since the latter converges to the 0-1 loss when $\kappa$ decreases to zero. 
With the same variance reduction and negative reward handling trick, we propose the \textit{FITR-Ramp} method to estimate $f_1$ by
\begin{equation}
\label{equ:FITR.Ramp}
\begin{split}
    &\widehat{f}_{1n} = \argmin_{f_1 \in \cH} 
    \frac{1}{n} \sum_{i = 1}^n \frac{\abs{R_{i1} - \bbE(R_{i1} | \bX_i)}}{\pi(A_i; \bX_i)} \cdot \\
    &\phi [A_i \sgn{R_{i1} - \bbE(R_{i1} | \bX_i)} f_1 (\bX_i)] 
    + \lambda_{1n} \norm{f_1}^2 \\
    &+ \frac{\mu_{1n}}{n} \sum_{i = 1}^n \sum_{k = 2}^K \Omega_{1k} \psi_{\kappa_{1n}} [f_1 (\bX_i) \widetilde{f}_k (\bX_i)].
\end{split}
\end{equation}
We allow $\kappa_{1n}$ to depend on the sample size $n$.

\paragraph{FITR-IntL.}
Alternatively, we can solve the optimization problem in (\ref{equ:fkn.hat}) using the following procedure. 
Notice that the disagreement between $f_1$ and $\widetilde{f}_k$ can be expressed with the disagreement between $A_i$ and $f_1$ since
\begin{equation*}
\begin{split}
    & \bbone \{f_1 (\bX_i) \widetilde{f}_k (\bX_i) < 0\} \\
    = & \bbone \{A_i f_1 (\bX_i) < 0\} \bbone \{A_i \widetilde{f}_k (\bX_i) > 0\} \\
    & + [1- \bbone \{A_i f_1 (\bX_i) < 0\}] \bbone \{A_i \widetilde{f}_k (\bX_i) < 0\} \\
    = & \bbone \{A_i f_1 (\bX_i) < 0\} \sgn{A_i \widetilde{f}_k (\bX_i)} + \bbone \{A_i \widetilde{f}_k (\bX_i) < 0\}.
\end{split}
\end{equation*}
The last term does not depend on $f_1$ given observed data.
Therefore, the problem in (\ref{equ:fkn.hat}) is equivalent to
        $\widehat{f}_{1n} = \argmin_{f_1 \in \cH} 
        \frac{1}{n} \sum_{i = 1}^n \frac{\widetilde{R}_{i1}}{\pi(A_i; \bX_i)} \bbone \{A_i f_1 (\bX_i) < 0\} 
        + \lambda_{1n} \norm{f_1}^2$,
where 
\[ \widetilde{R}_{i1} = R_{i1} + \mu_{1n} \pi(A_i; \bX_i) \sum_{k=2}^K \Omega_{1k} \sgn{A_i \widetilde{f}_k (\bX_i)} \]
is the pseudo outcome.
We then substitute the indicator function by the logistic loss and estimate $\widehat f_{1n}$ by
\begin{equation}
\label{equ:FITR.IntL}
    \begin{split}
        & \widehat{f}_{1n} = \argmin_{f_1 \in \cH} 
        \frac{1}{n} \sum_{i = 1}^n \frac{\absn{\widetilde{R}_{i1} - \bbE(\widetilde{R}_{i1} | \bX_i)}}{\pi(A_i; \bX_i)} \cdot \\
        & \phi[A_i \sgn{\widetilde{R}_{i1} - \bbE(\widetilde{R}_{i1} | \bX_i)} f_1 (\bX_i)] + \lambda_{1n} \norm{f_1}^2
    \end{split}
\end{equation}
with the same variance reduction and negative reward handling trick.
We refer to this optimization method as \textit{FITR-IntL}. It is worth noting that a similar procedure was originally proposed in \citeauthor{Intlearn} (in press) and \citet{qiu2018statistical} to integrate treatment rules from multiple studies for the same outcome without theoretical justifications, whereas our focus is on integrating multiple outcomes.

\paragraph{Implementation.}
In SepL, FITR-Ramp and FITR-IntL, the semi-norm for $f_1$ is usually chosen from a reproducing kernel Hilbert space (RKHS) associated with a real-valued kernel function $k: \cX \times \cX \to \bbR$. The choice of $k$ can be the linear kernel $k(\bx, \bx^{\prime})=\bx^T \bx^{\prime}$ which yields a linear decision function for $f_1$, or the Gaussian kernel $k(\bx, \bx^{\prime})=\exp (- \sigma_{1n}^2 \norm{\bx - \bx^{\prime}}_2^2)$ which yield a nonlinear decision function, where $\sigma_{1n}$ is a parameter depending on $n$. 
By the representer theorem, the minimizer for $f_1$ takes the form $f(\bX) = \sum_{i=1}^n \alpha_i k(\bX, \bX_i)$, so the optimization problems can be restricted to class
$ \cH := \{f: f(\bX) = \sum_{i=1}^n \alpha_i k(\bX, \bX_i), (\alpha_1, \dots, \alpha_n) \in \bbR^n \} $.
The function class $\cH$ is nonparametric when the RKHS with Gaussian kernel is used. When $\mu_{1n}, \kappa_{1n} \to 0$, $\sigma_{1n}^2 \to \infty$ as $n \to \infty$, this nonparametric method is asymptotically equivalent to unconstrained problems without penalties. Therefore, the learned FITR is asymptotically optimal even when the non-constant shift $E (R | \bX)$ and the fusion penalty are introduced.

For computation, FITR-Ramp can be solved by the difference of convex algorithm (DCA) \citep{le2018dc} or the Powell algorithm \citep{powell1964efficient,press2007numerical}.
DCA re-expresses the objective function as the difference between two convex functions, whereas the Powell algorithm is suitable for non-differentiable objective functions.
In contrast, FITR-IntL has a differentiable convex objective function so that it can be easily solved by gradient-based algorithms including variations of the gradient descent algorithm \citep{ruder2016overview} or the BFGS algorithm \citep{fletcher1987practical}.
The tuning parameters $\lambda_{1n}, \mu_{1n}, \kappa_{1n}$ can be selected by cross-validation.
The computation time of one replication when $n = 200$ is about 0.003 seconds for SepL, 0.003 seconds for FITR-IntL, 0.086 seconds for FITR-Ramp when using the linear kernel, 
and about 0.457 seconds for SepL, 0.356 seconds for FITR-IntL, 2.744 seconds for FITR-Ramp when using the Gaussian kernel.

\section{THEORETICAL RESULTS} \label{sec:theory}

In this section, we provide theoretical results about the agreement rate between $\widehat{f}_{1n}$ and $f_2^*$ in FITR-Ramp.
See Section~\ref{suppsec:sub.theory} in the supplementary material for the convergence rates of the value function $\cV_1 (\widehat{f}_{1n})$ and the misclassification rate $\bbP (\widehat{f}_{1n} f_1^* < 0)$. 

Without loss of generality, we assume that $\widetilde{f}_k: \cX \to \{1, -1\}$ is a binary mapping learned from a dataset of size $N_k$ for all $k = 2, \dots, K$, since every decision function $f$ can be transformed into a binary function by taking its sign.
In this section, we assume that the conditional expectation of $R_1$ has already been removed and that the signs have been flipped if the remaining reward is negative.
Then the problem becomes
\begin{equation*}
    \begin{split}
        \min_{f_1 \in \cH} 
        \frac{1}{n} \sum_{i = 1}^n \frac{R_{i1}}{\pi(A_i; \bX_i)} \phi [A_i f_1 (\bX_i)]
        + \lambda_{1n} \norm{f_1}^2 \\
        + \frac{\mu_{1n}}{n} \sum_{i = 1}^n \sum_{k = 2}^K \Omega_{1k} \psi_{\kappa_{1n}} [f_1 (\bX_i) \widetilde{f}_k (\bX_i)].
    \end{split}
\end{equation*}
We introduce the following assumptions to derive the convergence rate of the agreement rate.

\begin{asp} \label{asp:bounded.reward}
    Suppose that $0 \le R_1 \le c_1$ for all $R_1$ and for some constant $c_1 > 0$.
\end{asp}

Assumption~\ref{asp:bounded.reward} assumes that the primary outcome is nonnegative and bounded.
Furthermore, with $r_1^{(a)} (\bX) := \bbE(R_1 | A = a, \bX) \ge 0$, we define 
\begin{equation*}
    \begin{split}
        \eta(\bX) := 
        \begin{cases}
            \frac{r_1^{(1)} (\bX)}{\sum_{a \in \cA} r_1^{(a)} (\bX)}, & \text{if } \sum_{a \in \cA} r_1^{(a)} (\bX) > 0, \\
            \frac{1}{2}, & \text{if } r_1^{(a)} (\bX) = 0 \text{ for all } a \in \cA.
        \end{cases}
    \end{split}
\end{equation*}
The definition of $\eta(\bX)$ aligns with the probability $P(y = 1 | x)$ in classification literature, where $x$ is the predictor, and $y$ is the label.
Denote the regions $\cX_{-1} := \brce{\bx \in \cX: \eta(\bx) < 1/2}$, $\cX_{1} := \brce{\bx \in \cX: \eta(\bx) > 1/2}$ and 
$\cX_{0} := \brce{\bx \in \cX: \eta(\bx) = 1/2}$.
Finally, we define a distance function $\bx \mapsto \omega_{\bx}$ as 
    \begin{equation*}
        \omega_\bx := 
        \begin{cases}
            d(\bx, \cX_0 \cup \cX_1), & \text{if } \bx \in \cX_{-1}, \\
            d(\bx, \cX_0 \cup \cX_{-1}), & \text{if } \bx \in \cX_{1}, \\
            0, & \text{otherwise},
        \end{cases}
    \end{equation*}
    where $d(\bx, \cS)$ denotes the distance of $\bx$ to a set $\cS$ with respect to the Euclidean norm.

\begin{asp} \label{asp:geometric.noise}
    Assume that there exists a constant $q > 0$ for the distribution of $\bX$ such that
    \begin{equation} \label{equ.geometric.noise}
        \int_{\cX} \exp \prth{- \frac{\omega_{\bx}^2}{t}} \bbP_{\cX} (d \bx) \le C t^{q d / 2}
    \end{equation}
    for all $t > 0$ and some constant $C > 0$.
    We say $q = \infty$ if the inequality holds for all $q > 0$.
\end{asp}

Note that Assumption~\ref{asp:geometric.noise} is used to bound the approximation error when we estimate ITRs in an RKHS, and the logistic loss is used.
It is stronger than the geometric noise assumption \citep{steinwart2007fast} in the sense that $\abs{2 \eta(\bx) - 1}$ is not included in the left-hand side of (\ref{equ.geometric.noise}) and $\abs{2 \eta(\bx) - 1} \le 1$.
However, when $t \to 0$, we can still ensure that the left-hand side of (\ref{equ.geometric.noise}) goes to zero.

\begin{asp} \label{asp:conv.secondary}
    For any $k = 2, \dots, K$, assume that the estimator of the secondary outcome ITR $\widetilde{f}_k$ converges to $f_k^*$ in the sense that $\bbP(\widetilde{f}_k f_k^* < 0) \le \widetilde{\delta}_{kN_k}(\tau) $
    with probability at least $1 - e^{\tau}$,
    where $N_k$ is the sample size of the dataset for learning $\widetilde{f}_k$ and $\widetilde{\delta}_{kN_k}(\tau) = o(1)$ as $N_k \to \infty$.
\end{asp}

Assumption~\ref{asp:conv.secondary} is used to quantify the agreement rate between $\widetilde{f}_k$ and its corresponding optimal ITR $f_k^*$. It is a weak assumption that only requires $\widetilde{f}_k$ to converge to its true optimal ITR and can be satisfied by general methods for learning ITRs. We provide the convergence rate $\widetilde{\delta}_{k N_k}(\tau)$ in the remark of Corollary~\ref{thm:convergence.rate.accuracy} when $\widetilde{f}_k$ is learned by SepL.

We first focus on the case when $K=2$.
That is, there exists only one secondary treatment rule $\widetilde{f}_2$, which is already estimated from an external or the same dataset.
Define the first and second part in the loss function as
\begin{align*}
    \ell_1 \circ f_1 (\bZ) &:= \frac{R_1}{\pi (A | \bX)} \phi [A f_1 (\bX)] \text{ and} \\
    \ell_2 \circ f_1 (\bZ) &:= \mu_{1n} \Omega_{12} \psi_{\kappa_{1n}} [f_1 (\bX) \widetilde{f}_2 (\bX)].
\end{align*}
Let the risk based on the surrogate losses be 
\[ \cR (f_1) := \bbE \brck{\ell_1 \circ f_1 + \ell_2 \circ f_1}. \]
In the following inequalities, ``$\lesssim$'' indicates that the left-hand side is no larger than the right-hand side for all $n$ up to a universal constant. 

\begin{lem} \label{lem:surrogate.risk}
    Under Assumptions~\ref{asp:ignorability}-\ref{asp:positivity} and~\ref{asp:bounded.reward}-\ref{asp:geometric.noise},
    when $\mu_{1n}, \kappa_{1n} \to 0$, $\sigma_{1n}^2 \to \infty$, and $\kappa_n \le 1$ for all $n$,
    for any constant $\delta > 0$, $0 < \nu < 2$ and for all $\tau \ge 1$, we have 
    $\bbP (\cR (\widehat{f}_{1n}) - \cR (f_1^*) \lesssim \delta_{1n}(\tau)) \ge 1 - e^{-\tau}$, where
    \begin{equation} \label{equ:delta.n}
        \begin{split}
            \delta_{1n}(\tau) :=
            \lambda_{1n}^{-\frac{1}{2}} n^{-\frac{1}{2}} \brck{\sqrt{\tau} + \sigma_{1n}^{(1 - \nu/2) (1 + \delta) d} \gamma_{1n}^{\nu}} \\
            + \lambda_{1n} \sigma_{1n}^d + \gamma_{1n} (2d)^{q d / 2} \sigma_{1n}^{-q d},
        \end{split}
\end{equation}
    and 
    $ \gamma_{1n} := 1 + \mu_{1n} \kappa_{1n}^{-1}$.
\end{lem}

The first term on the right-hand side of (\ref{equ:delta.n}) is the estimation error, and the sum of the last two terms is the approximation error.
A larger class $\cH$ generally leads to a larger estimation error and a smaller approximation error, which corresponds to small penalty parameters $\lambda_{1n}$ and $\mu_{1n}$ with less restriction on the norm of the space or the agreement with the secondary outcome ITR.
The optimal rate that balances the estimation error and approximation error is 
\begin{equation} \label{equ:min.delta.n}
    \delta_{1n}(\tau) = \gamma_{1n}^{\frac{2q \nu + 2\Delta + 1}{3q + 2\Delta + 1}} n^{- \frac{q}{3q + 2\Delta + 1}},
\end{equation}
where $\Delta = (1 - \nu/2) (1 + \delta)$.
Note that when $\mu_{1n} = 0$, FITR degenerates to SepL learned with the logistic loss.
In this case, the optimal convergence rate is 
\begin{equation} \label{equ:min.delta.n0}
    \delta_{1n}^{(0)}(\tau) = n^{- \frac{q}{3q + 2\Delta + 1}}.
\end{equation}
The optimal tuning parameters are provided in Section~\ref{suppsec:sub.details.optimal.rates}.


Then we can present the convergence rates of the agreement rate $\bbP (\widehat{f}_{1n} f_2^* > 0)$.

\begin{thm} \label{thm:convergence.rate.disagree}
    Under Assumptions~\ref{asp:ignorability}-\ref{asp:positivity} and~\ref{asp:bounded.reward}-\ref{asp:conv.secondary}, 
    the agreement rate between $\widehat{f}_{1n}$ and $f_2^*$ satisfies
    \begin{equation} \label{equ:convergence.rate.disagree}
        \bbP (f_1^* f_2^* > 0) - \bbP (\widehat{f}_{1n} f_2^* > 0) \lesssim \frac{\delta_{1n}(\tau)}{\mu_{1n}} + \widetilde{\delta}_{2 N_2}(\tau)
    \end{equation}
    with probability at least $1 - 2 e^{\tau}$.
\end{thm}

Now we compare the agreement rate with or without the fusion penalty.
Note that for SepL,
\begin{equation} \label{equ:convergence.rate.disagree1}
    \begin{split}
        \bbP (f_1^* f_2^* > 0) - \bbP (\widehat{f}_{1n} f_2^* > 0)
        & \le \bbP (\widehat{f}_{1n} f_1^* < 0) \\
        & \lesssim n^{- \frac{\alpha}{2 - \alpha} \frac{q}{3q + 2\Delta + 1}}
    \end{split}
\end{equation}
with high probability under additional Assumptions~\ref{asp:Tsybakov.noise} and~\ref{asp:nonzero.rewardsum} (see details in Section~\ref{suppsec:sub.details.optimal.rates}).
On the other hand, for FITR-Ramp, (\ref{equ:convergence.rate.disagree}) and (\ref{equ:min.delta.n}) implies 
\begin{equation} \label{equ:convergence.rate.disagree2}
    \begin{split}
        & \bbP (f_1^* f_2^* > 0) - \bbP (\widehat{f}_{1n} f_2^* > 0) \\
        \lesssim & \mu_{1n}^{-1} \gamma_{1n}^{\frac{2q \nu + 2\Delta + 1}{3q + 2\Delta + 1}} n^{- \frac{q}{3q + 2\Delta + 1}} + \widetilde{\delta}_{2 N_2}(\tau)
    \end{split}
\end{equation}
with high probability.
The inequalities (\ref{equ:convergence.rate.disagree1}) and (\ref{equ:convergence.rate.disagree2})
suggest that SepL can learn the disagreement rate faster than FITR-Ramp when the data are fully separated by the decision boundary, i.e. when $\alpha = 1$.
However, FITR-Ramp has a faster convergence rate when
\begin{equation} \label{equ:compare.SepL.FITR}
    \begin{cases}
        \mu_{1n}^{3q - 2q \nu} \kappa_{1n}^{2q \nu + 2\Delta + 1} \gtrsim n^{- \frac{2 - 2 \alpha}{2 - \alpha} q}, &
        \text{if } \mu_{1n} \gtrsim \kappa_{1n}, \\
        \mu_{1n} \gtrsim n^{- \frac{2 - 2 \alpha}{2 - \alpha} \frac{q}{3q + 2\Delta + 1}}, &
        \text{otherwise,}
    \end{cases}
\end{equation}
and $\widetilde{f}_2$ is learned using SepL and $N_2 \gg n$.
The requirement of the sample size $N_2$ is consistent with the conclusion in transfer learning literature \citep{li2021transfer,tian2022transfer}, which also requires the sample size of auxiliary data to be much larger than the sample size of target data.
The relationship (\ref{equ:compare.SepL.FITR}) holds when, for example, $\alpha = 1/2, \nu = 3/2, \delta = 1, \Delta = 1/2, \min\{\mu_{1n}, \kappa_{1n}\} \gtrsim n^{- 2q / (9q+6)}$, and the sample size $N_2 \gtrsim n^3 \min\{\mu_{1n}, \kappa_{1n}\}^{(9q+6) / q} \gtrsim n$.
That is, the data are not fully separated, and the tuning parameters $\mu_{1n}, \kappa_{1n}$ are not decreasing too fast.
Besides, (\ref{equ:convergence.rate.disagree2}) does not rely on Assumptions~\ref{asp:Tsybakov.noise} and~\ref{asp:nonzero.rewardsum}.
The former is Tsybakov's noise assumption, and the latter assumes that for each patient, at least one treatment can yield a positive mean reward.

When $K \ge 2$, Theorem~\ref{thm:convergence.rate.disagree} can be generalized as follows.

\begin{thm} \label{thm:convergence.rate.K}
    Under Assumptions~\ref{asp:ignorability}-\ref{asp:positivity} and~\ref{asp:bounded.reward}-\ref{asp:conv.secondary}, 
    the agreement rate between $\widehat{f}_{1n}$ and any $f_k^*$ for $k\ge 2$ satisfies
    \begin{equation*}
    \label{equ:convergence.rate.disagree.K}
        \bbP (f_1^* f_k^* > 0) - \bbP (\widehat{f}_{1n} f_k^* > 0) \lesssim \frac{\delta_{1n}(\tau)}{\mu_{1n}} + \sum_{k=2}^K \widetilde{\delta}_{k N_k}(\tau)
    \end{equation*}
    with probability at least $1 - K e^{\tau}$.
\end{thm}

\section{SIMULATION STUDY} \label{sec:simulation}

In the simulation study, we learn FITR for the primary outcome using the proposed methods FITR-Ramp and FITR-IntL, comparing them with the baseline method SepL.
Suppose the ITR $\widetilde{f}_{k N_k}$ for the secondary outcome $R_k$ is estimated with SepL using an external dataset of sample size $N_k$.
Even though the parameter dimension of the Gaussian kernel in RKHS depends on the sample size, the values of $n$ and $N_k$ can be different for $\widehat{f}_{1n}$ and $\widetilde{f}_{k N_k}$
since the fusion penalty relies on secondary outcome ITRs only through the treatment suggestions.
The weight $\Omega_{1k}$ is chosen to be the Pearson correlation between $R_1$ and $R_k$.
Our experiments show that Spearman's rank correlation \citep{zar2014spearman} yields similar results.
Further implementation details can be found in Section~\ref{suppsec:sub.simulation.imple}.

\begin{figure}[!t]
    \centering
    \includegraphics[width=0.45\textwidth]{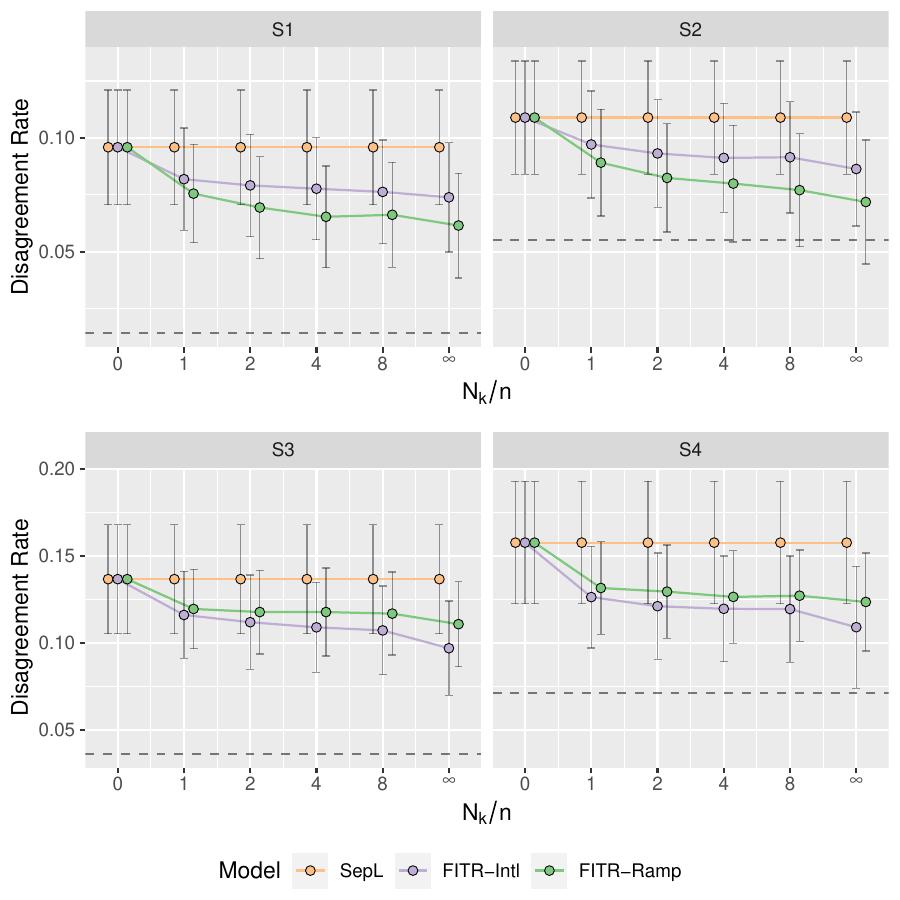}
    \caption{The mean and standard deviation of disagreement rate between FITR $\widehat{f}_{1n}$ learned using SepL, FITR-Ramp or FITR-IntL, and the true secondary outcome ITR $f_2^*$. The dashed line represents the true disagreement rate $\bbP(f_1^* f_2^* < 0)$.}
    \label{fig:K2.disagree}
\end{figure}

\paragraph{Simulation Settings.}
We assume the data are collected from a randomized controlled trial and $\pi (1 ; \bX_i) = \pi (-1 ; \bX_i) = 0.5$ for all $i = 1, \dots, n$.
The $k$th outcome is defined as $R_{ik} = m_k (\bX_i) + T_k (\bX_i, A_i) + \epsilon_k (\bX_i, A_i)$, where $m_k$ is the main effect, $T_k$ is the interaction effect between the covariates and the treatment, and $\epsilon_k$ is the noise term.
By choosing different $T_k$, we allow both linear and nonlinear treatment rules.
We let the dimension of covariates be $d = 10$ (see details in Section~\ref{suppsec:sub.simulation.data}).
The simulation process is repeated 400 times under each scenario.

The sample size of the external dataset $N_k$ is set to $r \cdot n$ for $k \ge 2$, with $r$ taking values from the set $\{0, 1, 2, 4, 8, \infty\}$.
In the case of $r = 0$, we do not estimate $\widetilde{f}_{k N_k}$, and FITR degenerates to SepL.
When $r = \infty$, we essentially have $\widetilde{f}_{k N_k} = f^*_{k}$, which represents the scenario where the true ITR $f^*_{k}$ is known.

We first assess performance with $K = 2$ and $n = 200$.
Consider the following four scenarios.
In all scenarios, the main effects are defined to be nonlinear functions of $\bX$,
\begin{align*}
    m_1 (\bX) &= 1 + 2 X_1 + X_2^2 + X_1 X_2, \\
    m_2 (\bX) &= 1 + 2 X_1^2 + 1.5 X_2 + 0.5 X_1 X_2.
\end{align*}
The interaction terms are defined as follows:
\begin{itemize}
    \item In linear scenarios S1 and S2, \\
    $T_1 (\bX, A) = 0.5 A (0.2 - X_1 - 2 X_2)$,\\
    $T_2 (\bX, A) = 0.8 A (0.2 - X_1 - \gamma_1 X_2)$,\\
    where $\gamma_1 = 1.8$ in S1 and $\gamma_1 = 1.4$ in S2.
    \item In nonlinear scenarios S3 and S4, \\
    $T_1 (\bX, A) = 1.0 A (-2.2 - e^{X_1} - e^{X_2})$, \\
    $T_2 (\bX, A) = 1.5 A (-\gamma_2 - e^{X_1} - e^{X_2})$,\\
    where $\gamma_2 = 2.3$ in S3 and $\gamma_2 = 2.4$ in S4.
\end{itemize}
The true agreement rate $\bbP(f_1^* f_2^* > 0)$ is 98.6\% for S1, 94.5\% for S2, 96.4\% for S3, and 92.8\% for S4.
Table~\ref{tbl:true.opt.K2} presents the true optimal value functions for each scenario.

\paragraph{Simulation Results.}
Our experiments show that the linear kernel performs better in linear scenarios, while the Gaussian kernel is more effective in nonlinear scenarios. 
We present results using only the most suitable kernel for each scenario.

The average disagreement rate $\bbP(\widehat{f}_{1n} f_2^* < 0)$ across all replications, along with its standard deviation, is plotted in Figure~\ref{fig:K2.disagree} for different $N_k$.
As expected, the disagreement rate remains unchanged as $N_k$ increases if $\widehat{f}_{1n}$ is learned by SepL.
In contrast, the fusion penalty's effect on reducing the disagreement rate becomes more significant with increasing $N_k$.
Specifically, FITR-Ramp is better in linear scenarios, while FITR-IntL outperforms it in nonlinear scenarios. 
This distinction may be attributed to the fact that FITR-IntL imposes a heavier penalty on the disagreement rate when the parameter dimension is high.
The improvement in the agreement rate slows down after $N_k/n \ge 2$, indicating that a moderately small external dataset is sufficient for learning an FITR with comparable performance to using the true $f_2^*$.
As the difference between outcomes increases, the disagreement rate increases for all methods, but an improvement with FITR is still observed.

\begin{figure}[!t]
    \centering
    \begin{subfigure}{0.45\textwidth}
        \centering
        \includegraphics[width=\textwidth]{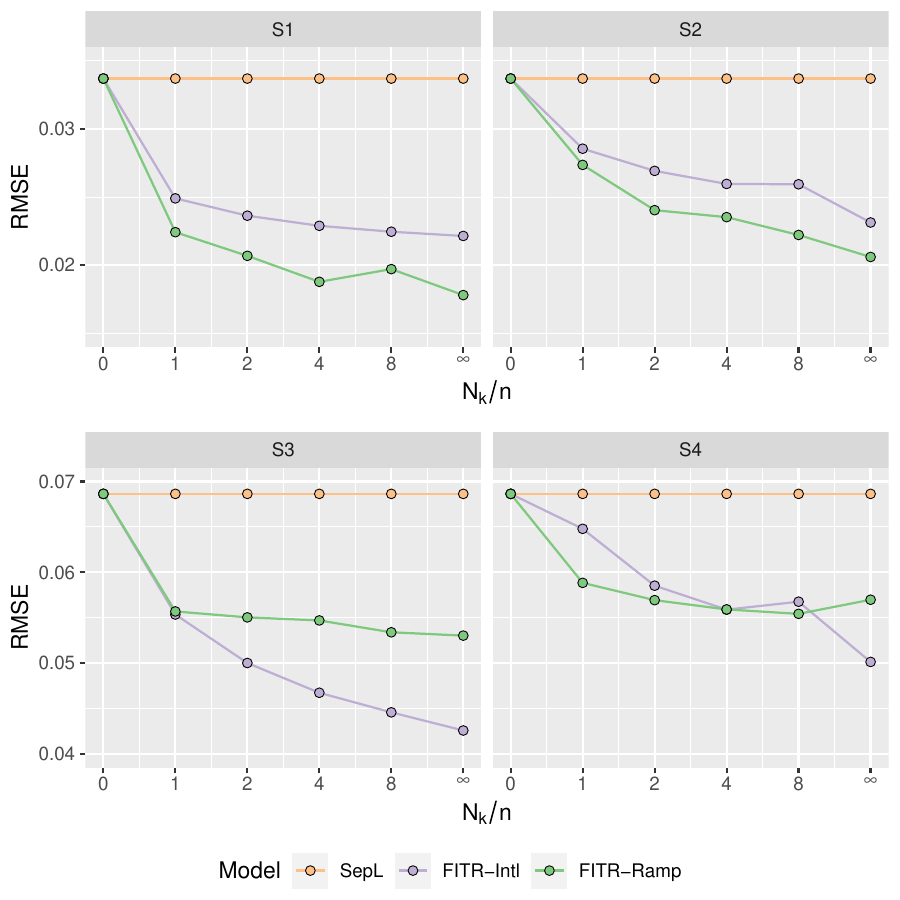}
        \caption{RMSE}
        \label{subfig:rmse}
    \end{subfigure}
    \hfill
    \begin{subfigure}{0.45\textwidth}
        \centering
        \includegraphics[width=\textwidth]{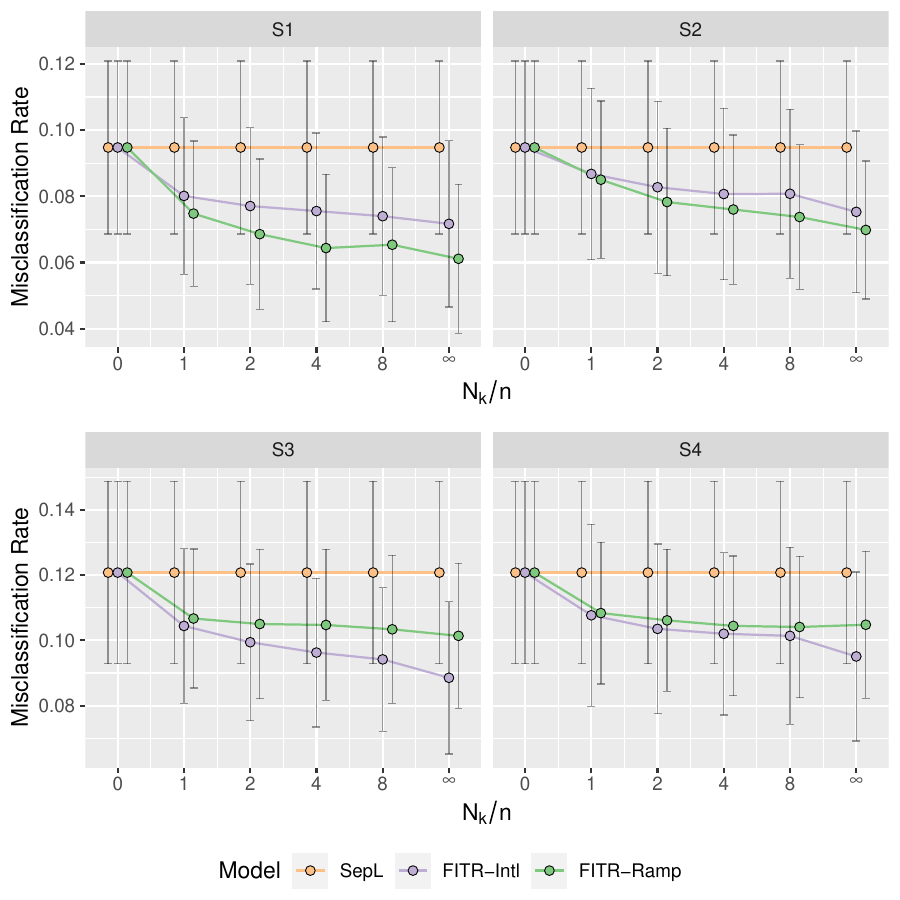}
        \caption{Misclassification rate}
        \label{subfig:accuracy}
    \end{subfigure}
    \caption{(a) The RMSE of value functions and (b) the mean and standard deviation of misclassification rate.}
    \label{fig:K2.rmse.accuracy}
\end{figure}

In Figure~\ref{fig:K2.rmse.accuracy}, we evaluate FITR's ability to generate treatment suggestions for the primary outcome. 
We present two metrics: (a) the root mean square error (RMSE) of the value function $\cV_1 (\widehat{f}_{1n})$ in comparison to the optimal value function $\cV_1 (f^*_{1})$, and
(b) the average misclassification rate $\bbP (\widehat{f}_{1n} f_1^* < 0)$ along with its standard deviation.
As an additional advantage of the proposed method, the primary outcome is actually improved by the FITR with the help of the secondary outcome.
Similarly, FITR-Ramp in S1, S2 and FITR-IntL in S3, S4 observe the minimum RMSE and misclassification rate among all methods.
When the discrepancy between outcomes increases, the performance of SepL remains unaffected, but the improvement by FITR-Ramp and FITR-IntL is reduced.

Additional results for different values of $K$ and $n$ are available in Section~\ref{suppsec:sub.simulation}.
When $K = 2$ and $n = 100$ in scenarios S3 and S4, and the Gaussian kernel is used, FITR-Ramp instead of FITR-IntL yields the minimum RMSE of the value function.
This indicates that FITR-IntL may be unstable when the sample size is small, and the parameter dimension is high.
Furthermore, in cases where $K = 3$, the disagreement rate, RMSE, and misclassification rate further decrease.
The FITR performs better for the primary outcome as long as either $R_2$ or $R_3$ is close to $R_1$, even if $f^*_2$ and $f^*_3$ differ from $f^*_1$ in opposite directions, suggesting the advantages of incorporating multiple secondary outcomes.
In practice, cross-validation can be used to choose the best model and kernel.

\paragraph{Sensitivity Analysis.}
To demonstrate the influence of the similarity between outcomes on the estimated FITRs, we fix the primary outcome and vary the secondary outcome when $K = 2$.
Specifically, we use the same main effect and noise term as in S1 and let the interaction effect be
\begin{align*}
    T_1 (\bX, A) &= 0.5 A (0.2 - X_1 - 2 X_2), \\
    T_2 (\bX, A) &= 0.8 A (0.2 - X_1 - 2 \rho X_2),
\end{align*}
where $\rho$ controls the similarity between $R_1$ and $R_2$.
In Table~\ref{tbl:sensitivity}, we summarize different values of $\rho$ and their corresponding true agreement rates.
The performance metrics, including the disagreement rate, RMSE, and misclassification rate, are presented when $n = 200$ and $N_2 = 4n$.
The table shows that the fusion penalty consistently increases the agreement rate between $\widehat{f}_{1n}$ and $f_2^*$.
However, the value function and accuracy of FITRs exceed SepL only when $\rho \ge 0.5$.
Although $\mu_{1n}$ tuned by cross-validation should ideally be zero when the outcomes are substantially different, practical constraints such as a small sample size may result in $\mu_{1n} > 0$, leading to decreased value functions.
This suggests that the true agreement rate should be no less than 87\% in this scenario for the fusion penalty to have a positive impact on the treatment suggestions for the primary outcome.

\section{REAL DATA ANALYSIS} \label{sec:realdata}

We apply our proposed methods to a study of major depressive disorder (MDD)  \citep{trivedi2016establishing} which randomized MDD patients to serotonin selective reuptake inhibitor sertraline (SERT) or placebo (PBO).

\begin{table}[!t]
    \small
    \centering
    \caption{
    The upper panel shows the agreement rates between the FITR $\widehat{f}_{\text{QIDS}}$ of the primary outcome QIDS-change, learned from SepL, FITR-IntL or FITR-Ramp, and the secondary outcome ITRs $\widetilde{f}_{\text{CGI}}$, $\widetilde{f}_{\text{SAS}}$.
    The lower panel shows the estimated value functions of the OSFA strategy, SepL, FITR-IntL, and FITR-Ramp for the primary outcome. }
    \setlength{\tabcolsep}{5pt}
    \begin{tabular}{ccccccc}
        \toprule
        \multicolumn{7}{c}{\makecell{Agreement rates between ITRs\\of primary and secondary outcomes}}\\ 
        \cmidrule(lr){1-7}
        \multicolumn{3}{c}{Secondary ITR} 
        & & $\widetilde{f}_{\text{CGI}}$ & & $\widetilde{f}_{\text{SAS}}$ \\
        \cmidrule(lr){1-7}
        \multicolumn{3}{c}{\multirow{3}{*}{\makecell{Primary ITR\\learned from}}}
            & SepL      & 0.807 & & 0.694 \\ 
        & & & FITR-IntL & 0.903 & & 0.737 \\ 
        & & & FITR-Ramp & 0.887 & & 0.720 \\ 
        \bottomrule
        \toprule
        \multicolumn{7}{c}{Value functions from 400 replicates}\\ 
        \cmidrule(lr){1-7}
          \makecell{All\\SERT} 
        &
        & \makecell{All\\PBO} 
        & \makecell{SepL\\\;}
        & \makecell{FITR-\\IntL} 
        &
        & \makecell{FITR-\\Ramp} \\
        \cmidrule(lr){1-7}
          8.000 
        &
        & 6.438 
        & \makecell{8.869\\(0.286)} 
        & \makecell{9.082\\(0.317)} 
        &
        & \makecell{9.015\\(0.327)} \\ 
        \bottomrule
    \end{tabular}
    \label{tbl:EMBARC.value}
\end{table}

\paragraph{Description of the Dataset.}
The metric for depressive symptoms is the Quick Inventory of Depressive Symptomatology (QIDS) score, with a lower score indicating better relief of symptoms.
Seven covariates are used for learning ITRs, including the NEO-Five Factor Inventory score (NEO), Flanker Interference Accuracy score (Flanker), sex, age, education years, Edinburgh Handedness Inventory (EHI) score, and the QIDS score at the baseline (see details in Section~\ref{suppsec:sub.realdata}).

The primary outcome is the reduction in QIDS at the end of the treatment phase compared to the baseline (QIDS-change), with a larger positive value indicating a more desirable improvement in symptoms. 
In addition to the primary outcome, we examine two secondary outcomes. 
One is the clinician-assessed Clinical Global Improvement scale (CGI) as an overall assessment of treatment effect, and a smaller score suggests better improvement \citep{busner2007clinical}.
The other is the Social Adjustment Scale (SAS), which evaluates the impact of a person's mental health difficulties, with a higher score indicating greater impairment. 
We consider the QIDS change and the negative values of CGI and SAS as the three outcomes. 
The dataset contains 186 patients with complete information.
Further details can be found in Section~\ref{suppsec:sub.realdata}.

\paragraph{Analysis Results.}
The weights $\Omega_{1k}$ for $k = 2,3$ are set by the Pearson correlation between the primary and secondary outcomes, which equals 0.72 and 0.19 for CGI and SAS, respectively.
Our experiments show that the linear kernel outperforms the Gaussian kernel on this dataset, so we present the results of the linear kernel. 
We examine the agreement rates between the ITRs of the primary outcome and secondary outcomes in the upper panel of Table~\ref{tbl:EMBARC.value}.
Since we do not have access to the true optimal secondary outcome ITR, the agreement rate is calculated based on the ITRs $\widetilde{f}_{\text{CGI}}$ and $\widetilde{f}_{\text{SAS}}$ estimated with SepL.
The table suggests that FITR-IntL or FITR-Ramp increases the agreement rate compared to SepL, especially between QIDS-change and CGI.

We also assess the value functions of the primary outcome for FITR-IntL, FITR-Ramp, SepL, and the one-size-for-all (OSFA) strategy (see the lower panel of Table~\ref{tbl:EMBARC.value}). 
The value function $\cV_1 (f_1)$ is estimated with the normalized inverse probability weighted estimator 
\begin{equation} \label{equ:test.value}
    \frac{\sum_{i=1}^n R_{i1} \bbone [A_i = \sgn{f_k (\bX_i)}] / \pi_i (A_i; \bX_i)}{\sum_{i=1}^n \bbone [A_i = \sgn{f_k (\bX_i)}] / \pi_i (A_i; \bX_i)},
\end{equation}
which is a consistent estimator.
The mean and standard deviation of the value functions across 400 replications are reported. 
Due to the limited sample size, we use 5-fold cross-validation in each replication, averaging the estimated value function across 5 validation sets to obtain the estimate for the current replication.
For OSFA, we directly calculate the mean reward on the subpopulation with the desired treatment, which is equivalent to (\ref{equ:test.value}) since $\pi_i = 0.5$ for all $i$.
The tuning parameters are selected using 4-fold cross-validation.
The results suggest that compared to OSFA and SepL, FITR-IntL and FITR-Ramp can improve the value functions for QIDS-change when assisted by the other secondary outcomes.
FITR-IntL has the best performance, and it increases the value function of SepL by approximately one standard deviation.

To compare the learned ITR using different methods, we present the coefficients of each variable in Table~\ref{tbl:EMBARC.itr}.
We observe that $\widehat{f}_{\text{QIDS}}$ and $\widetilde{f}_{\text{CGI}}$, when both estimated using SepL without the fusion penalty, share the same signs of the coefficients.
This indicates that QIDS-change and negative CGI indeed need similar treatments.
In addition, for QIDS-change, the signs of the coefficients in SepL, FITR-Ramp, and FITR-IntL are all consistent, suggesting that the fusion penalty does not change the ITR significantly from SepL.

\section{DISCUSSION} \label{sec:discussion}

In this work, we have proposed a method to optimize the primary outcome while also approximating the optimal secondary outcome ITRs as closely as possible.
A fusion penalty is introduced to encourage the ITRs to generate similar treatment recommendations. 
We demonstrate theoretically and numerically that the agreement rate between the FITR for the primary outcome and the ITR for the secondary outcome converges faster to its true value than SepL.
Moreover, the simulation and real data studies show that the ITR learned by FITR-Ramp and FITR-IntL have a higher value function and accuracy than SepL when the true optimal ITRs are closely aligned.

The proposed fusion penalty is a general method that can be easily extended to any variant of outcome-weighted learning \citep{zhao2015new,chen2016personalized,gao2022non,ma2023learning},
a classification-based method to estimate an ITR as the sign of some decision function.
There are several promising directions that are worth further research.
A possible extension is to let the tuning parameters $\lambda$ and $\mu$ depend on the covariates $\bX$ since the fusion level should vary if the optimal ITRs for the primary and secondary outcomes differ substantially for specific patients.
Another interesting direction is to explore a minimax bound for the convergence rate of the agreement rate. Such a bound might be tighter than the current results presented in Section~\ref{sec:theory}.
Nevertheless, the upper bounds for SepL and FITR are currently derived under the same conditions, allowing for a meaningful comparison between them.

\subsubsection*{Acknowledgements}
The authors express their gratitude to the anonymous reviewers for their insightful comments and valuable suggestions. 
This research was supported in part by US NIH grants GM124104, NS073671, and MH123487.

\bibliography{bibfile}

\section*{Checklist}



\begin{enumerate}

    \item For all models and algorithms presented, check if you include:
    \begin{enumerate}
        \item A clear description of the mathematical setting, assumptions, algorithm, and/or model. Yes, please see Sections~\ref{sec:methodology} and~\ref{sec:theory}.
        \item An analysis of the properties and complexity (time, space, sample size) of any algorithm. Yes, please see the sample size discussion in Sections~\ref{sec:simulation} and~\ref{suppsec:sub.simulation}, and see the computation time in Section~\ref{sec:sub.FITR.secondary}.
        \item (Optional) Anonymized source code, with specification of all dependencies, including external libraries. Yes, please see the supplementary material.
    \end{enumerate}

    \item For any theoretical claim, check if you include:
    \begin{enumerate}
        \item Statements of the full set of assumptions of all theoretical results. Yes, please see Sections~\ref{sec:theory} and~\ref{suppsec:sub.theory}.
        \item Complete proofs of all theoretical results. Yes, please see Sections~\ref{suppsec:sub.details.optimal.rates}-\ref{suppsec:sub.proof.K}.
        \item Clear explanations of any assumptions. Yes, please see Sections~\ref{sec:theory} and~\ref{suppsec:sub.theory}.
    \end{enumerate}

    \item For all figures and tables that present empirical results, check if you include:
    \begin{enumerate}
        \item The code, data, and instructions needed to reproduce the main experimental results (either in the supplemental material or as a URL). Yes, please see the code in the supplementary material and the simulation settings in Sections~\ref{sec:simulation} and~\ref{suppsec:sub.simulation}.
        \item All the training details (e.g., data splits, hyperparameters, how they were chosen). Yes, please see Sections~\ref{sec:sub.FITR.secondary} and~\ref{suppsec:sub.simulation.imple}.
        \item A clear definition of the specific measure or statistics and error bars (e.g., with respect to the random seed after running experiments multiple times). Yes, please see Sections~\ref{sec:simulation} and~\ref{suppsec:sub.simulation}.
        \item A description of the computing infrastructure used. (e.g., type of GPUs, internal cluster, or cloud provider). We only used CPU in our numerical experiments.
    \end{enumerate}

    \item If you are using existing assets (e.g., code, data, models) or curating/releasing new assets, check if you include:
    \begin{enumerate}
        \item Citations of the creator If your work uses existing assets. Yes, please see Section~\ref{sec:realdata}.
        \item The license information of the assets, if applicable. Not Applicable.
        \item New assets either in the supplemental material or as a URL, if applicable. Not Applicable.
        \item Information about consent from data providers/curators. We have signed the data use agreement with the NIMH to obtain the real data.
        \item Discussion of sensible content if applicable, e.g., personally identifiable information or offensive content. The real data that we use do not contain personally identifiable information or other sensible content.
    \end{enumerate}

    \item If you used crowdsourcing or conducted research with human subjects, check if you include:
    \begin{enumerate}
        \item The full text of instructions given to participants and screenshots. Not Applicable.
        \item Descriptions of potential participant risks, with links to Institutional Review Board (IRB) approvals if applicable. Not Applicable.
        \item The estimated hourly wage paid to participants and the total amount spent on participant compensation. Not Applicable.
    \end{enumerate}

\end{enumerate}

\end{document}


%
\runningtitle{Fusing Individualized Treatment Rules Using Secondary Outcomes}

%

\onecolumn
\aistatstitle{Fusing Individualized Treatment Rules Using Secondary Outcomes: \\
Supplementary Materials}

In the supplementary material, we provide additional implementation details and experiment results in the simulation study and real data analysis in Section~\ref{suppsec:simu}.
In addition, we provide additional theoretical results and their proof in Section~\ref{suppsec:theo}.

\appendix

\section{ADDITIONAL DETAILS AND EXPERIMENTS FOR SECTIONS~\ref{sec:simulation} AND~\ref{sec:realdata}} \label{suppsec:simu}

In this section, we provide some details about implementation and experiments in the simulation study of Section~\ref{sec:simulation} and the real data analysis of Section~\ref{sec:realdata}.

\subsection{Implementation Details} \label{suppsec:sub.simulation.imple}

To find the optimal solutions for FITR-IntL and FITR-Ramp, we use the \texttt{scipy} package in \texttt{Python}.
Our experiments show that the Powell algorithm generally produces better optimization results than DCA for FITR-Ramp, so we use the function \texttt{minimize(method=`Powell')} to minimize the loss function.
For FITR-IntL, we use the function \texttt{minimize(method=`BFGS')} with closed-form gradients specified for optimization.

The tuning parameter $\lambda_{1n}$ is first chosen with cross-validation when estimating the ITR using SepL for each reward $R_1, \dots, R_K$.
Subsequently, the parameter $\mu_{1n}$ in FITR-IntL or the parameters $\mu_{1n}$ and  $\kappa_{1n}$ in FITR-Ramp are tuned simultaneously with cross-validation while $\lambda_{1n}$ is kept fixed.
The parameter $\sigma_{1n}$ in the Gaussian kernel is chosen as the median of the distances between all covariate pairs, which is a common heuristic when the sample size is not too large \citep{garreau2017large}.

\subsection{Details for Data Generation and Value Function Estimation in Section~\ref{sec:simulation}} \label{suppsec:sub.simulation.data}

The first two covariates are important variables that affect the outcomes and are generated as $X_{ij} \overset{i.i.d.}{\sim} Unif(-1, 1)$ for all $i = 1, \dots, n$ and $j = 1, 2$.
Let $X_{i3} = 0.8 X_{i3}^{\prime} + X_{i1}$, where $X_{i3}^{\prime} \overset{i.i.d.}{\sim} Unif(-1, 1)$ so that $X_{i3}$ is correlated with $X_{i1}$.
The remaining covariates are independently generated as $X_{ij} \overset{i.i.d.}{\sim} Unif(-1, 1)$ for all $j = 4, \dots, d$.
Each patient's noise variable $(\epsilon_1, \dots, \epsilon_K)$ follows a mean-zero multivariate normal distribution, where the covariance matrix has 0.2 on its diagonal and 0.1 on its off-diagonal entries.

The value functions are numerically calculated from an independent test set of size 100,000.
The disagreement rate between two decision functions $\widehat{f}_{1n}$ and $f^*_k$ is estimated by averaging $\bbone \{\widehat{f}_{1n} f^*_k < 0\}$ on this test set.
The optimal value $\cV_k^*$ is obtained by averaging the rewards of all patients in the test set when the treatment is taken as $\argmax_{A} T_k (\bX, A)$ for the $k$th outcome.
The value function of each estimated ITR is obtained by following the corresponding treatment rule.
The misclassification rate of $\widehat{f}_{1n}$ is estimated by averaging $\bbone \{\widehat{f}_{1n} f^*_1 < 0\}$ on the test set.

\subsection{Additional Experiments for Estimating FITR in Section~\ref{sec:simulation}} \label{suppsec:sub.simulation}

For the scenarios considered in Section~\ref{sec:simulation}, we present their true optimal values in Table~\ref{tbl:true.opt.K2}.

\begin{table}[H]
    \centering
    \caption{True optimal values in S1-S4 when $K = 2$.}
    \begin{tabular}{cccccccc}
        \toprule
        \multicolumn{2}{c}{S1} & 
        \multicolumn{2}{c}{S2} & 
        \multicolumn{2}{c}{S3} & 
        \multicolumn{2}{c}{S4} \\
        \cmidrule(lr){1-2}
        \cmidrule(lr){3-4}
        \cmidrule(lr){5-6}
        \cmidrule(lr){7-8}
        $\cV^*_1$ & $\cV^*_2$ & 
        $\cV^*_1$ & $\cV^*_2$ & 
        $\cV^*_1$ & $\cV^*_2$ & 
        $\cV^*_1$ & $\cV^*_2$ \\
        1.89 & 2.47 & 
        1.89 & 2.33 & 
        2.12 & 2.82 & 
        2.12 & 2.83 \\
        \bottomrule
    \end{tabular}
    \label{tbl:true.opt.K2}
\end{table}

\begin{figure}[p]
    \centering
    \includegraphics[width=0.45\textwidth]{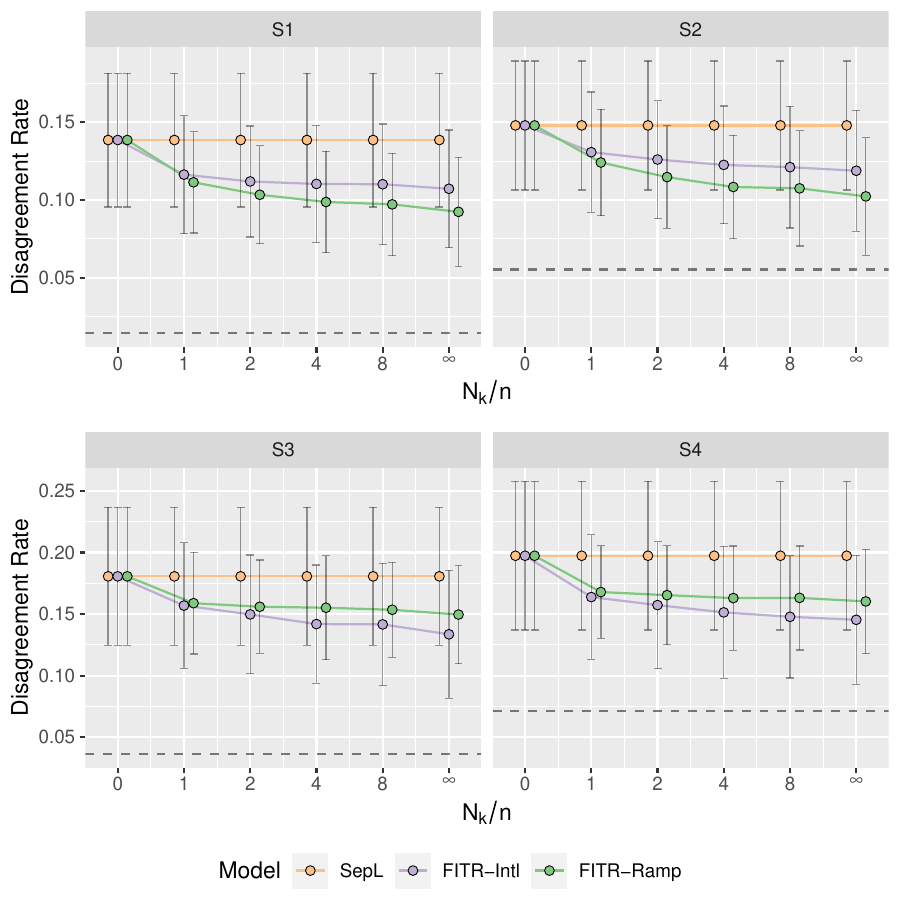}
    \caption{The mean and standard deviation of disagreement rate between FITR $\widehat{f}_{1n}$ learned using SepL, FITR-Ramp or FITR-IntL, and the true secondary outcome ITR $f_2^*$ when $K = 2$ and $n = 100$. The dashed line represents the true disagreement rate $\bbP(f_1^* f_2^* < 0)$.}
    \label{fig:K2.disagree.n100}
\end{figure}

\begin{figure}[p]
    \centering
    \begin{subfigure}{0.45\textwidth}
        \centering
        \includegraphics[width=\textwidth]{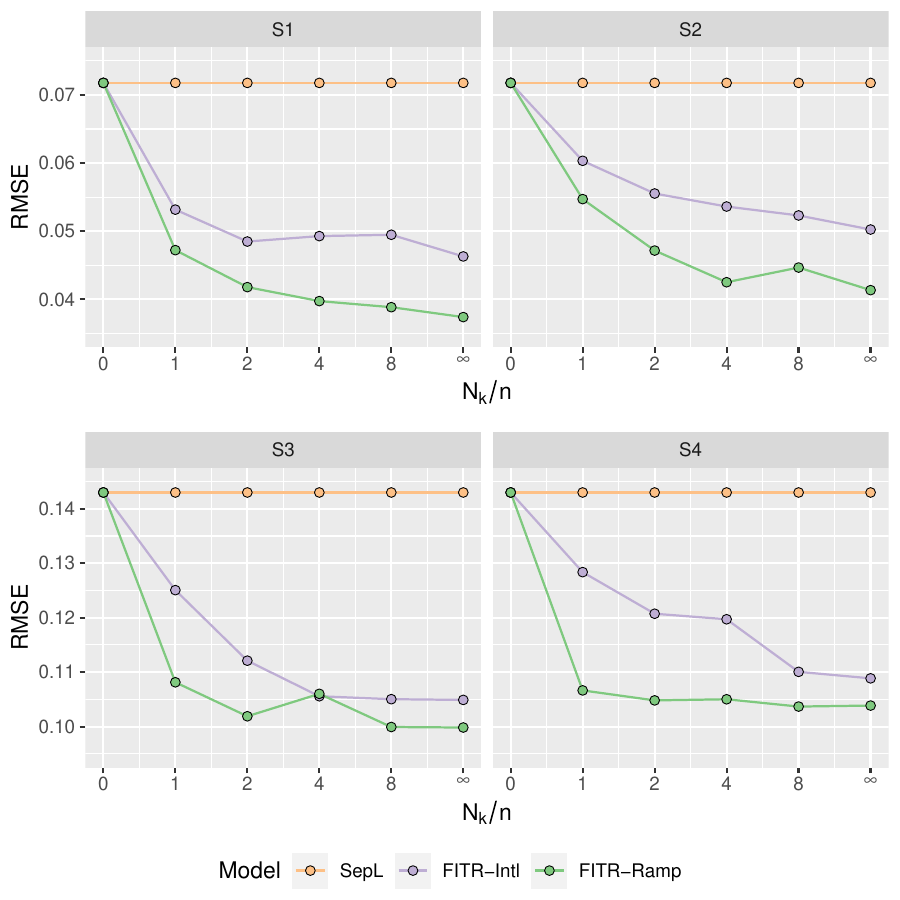}
        \caption{RMSE}
    \end{subfigure}
    \hfill
    \begin{subfigure}{0.45\textwidth}
        \centering
        \includegraphics[width=\textwidth]{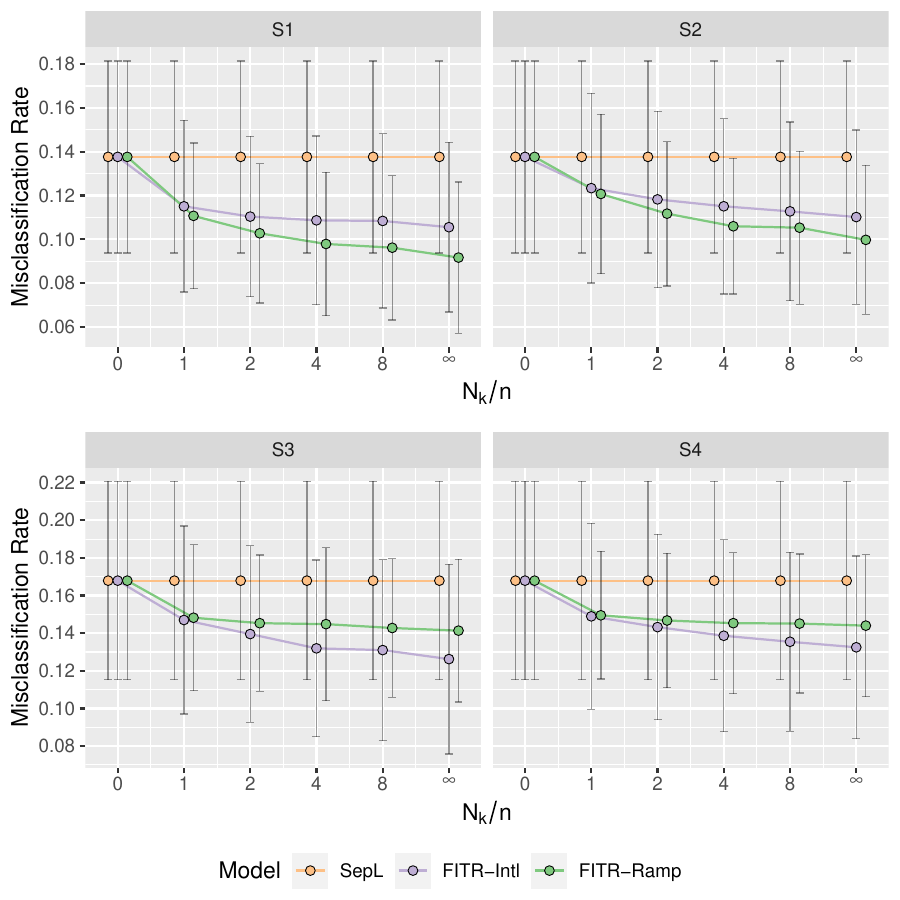}
        \caption{Accuracy}
    \end{subfigure}
    \caption{(a) The RMSE of value functions and (b) the mean and standard deviation of accuracy when FITR is learned using SepL, FITR-Ramp or FITR-IntL when $K = 2$ and $n = 100$.}
    \label{fig:K2.rmse.accuracy.n100}
\end{figure}

In addition to the results in Section~\ref{sec:simulation} when $n = 200$, we also present the case when $n = 100$ in Figures~\ref{fig:K2.disagree.n100} and ~\ref{fig:K2.rmse.accuracy.n100}.
The general conclusions are similar as before, but as the sample size decreases, the disagreement rate, RMSE and misclassification rate increases.
Besides, FITR-Ramp has smaller RMSE compared to FITR-IntL in S3 and S4, which is different from the case when $n = 200$. 
This may be due to the reason that FITR-IntL is unstable when the sample size is small and the variance is large.

We also explore the case where we have $K = 3$ outcomes.
Consider the following four scenarios.
In all scenarios, the main effects are
\begin{align*}
    m_1 (\bX) &= 1 + 2 X_1 + X_2^2 + X_1 X_2, \\
    m_2 (\bX) &= 1 + 2 X_1^2 + 1.5 X_2 + 0.5 X_1 X_2, \\
    m_3 (\bX) &= 1 + X_1 + X_2.
\end{align*}
The noise terms are the same as that in Section~\ref{suppsec:sub.simulation.data}.
The interaction terms are defined as follows:
\begin{itemize}
    \item In linear scenarios S5 and S6,
    \begin{align*}
        T_1 (\bX, A) &= 0.5 A (0.2 - X_1 - 2 X_2),\\
        T_2 (\bX, A) &= 0.8 A (0.2 - X_1 - \gamma_1 X_2),\\
        T_3 (\bX, A) &= 0.6 A (0.2 - X_1 - 2.2 X_2),
    \end{align*}
    where $\gamma_1 = 1.8$ in S5 and $\gamma_1 = 1.4$ in S6.
    \item In nonlinear scenarios S7 and S8, 
    \begin{align*}
        T_1 (\bX, A) &= 1.0 A (-2.2 - e^{X_1} - e^{X_2}), \\
        T_2 (\bX, A) &= 1.5 A (-\gamma_2 - e^{X_1} - e^{X_2}),\\
        T_3 (\bX, A) &= 1.5 A (-2.1 - e^{X_1} - e^{X_2}), 
    \end{align*}
    where $\gamma_2 = 2.3$ in S7 and $\gamma_2 = 2.4$ in S8.
\end{itemize}
The first two outcomes are the same as in Section~\ref{sec:simulation} when $K = 2$.
The true optimal values of each scenario is summarized in Table~\ref{tbl:true.opt.K3}.

\begin{table}[!ht]
    \centering
    \caption{True optimal values of S3 and S4 when $K = 3$.}
    \begin{tabular}{cccccccccccc}
        \toprule
        \multicolumn{3}{c}{S1} & 
        \multicolumn{3}{c}{S2} & 
        \multicolumn{3}{c}{S3} & 
        \multicolumn{3}{c}{S4} \\
        \cmidrule(lr){1-3}
        \cmidrule(lr){4-6}
        \cmidrule(lr){7-9}
        \cmidrule(lr){10-12}
        $\cV^*_1$ & $\cV^*_2$ & $\cV^*_3$ & 
        $\cV^*_1$ & $\cV^*_2$ & $\cV^*_3$ & 
        $\cV^*_1$ & $\cV^*_2$ & $\cV^*_3$ & 
        $\cV^*_1$ & $\cV^*_2$ & $\cV^*_3$ \\
        1.89 & 2.47 & 1.72 & 
        1.89 & 2.33 & 1.72 & 
        2.12 & 2.82 & 2.18 & 
        2.12 & 2.83 & 2.18 \\
        \bottomrule
    \end{tabular}
    \label{tbl:true.opt.K3}
\end{table}

\begin{figure}[p]
    \centering
    \includegraphics[width=0.8\textwidth]{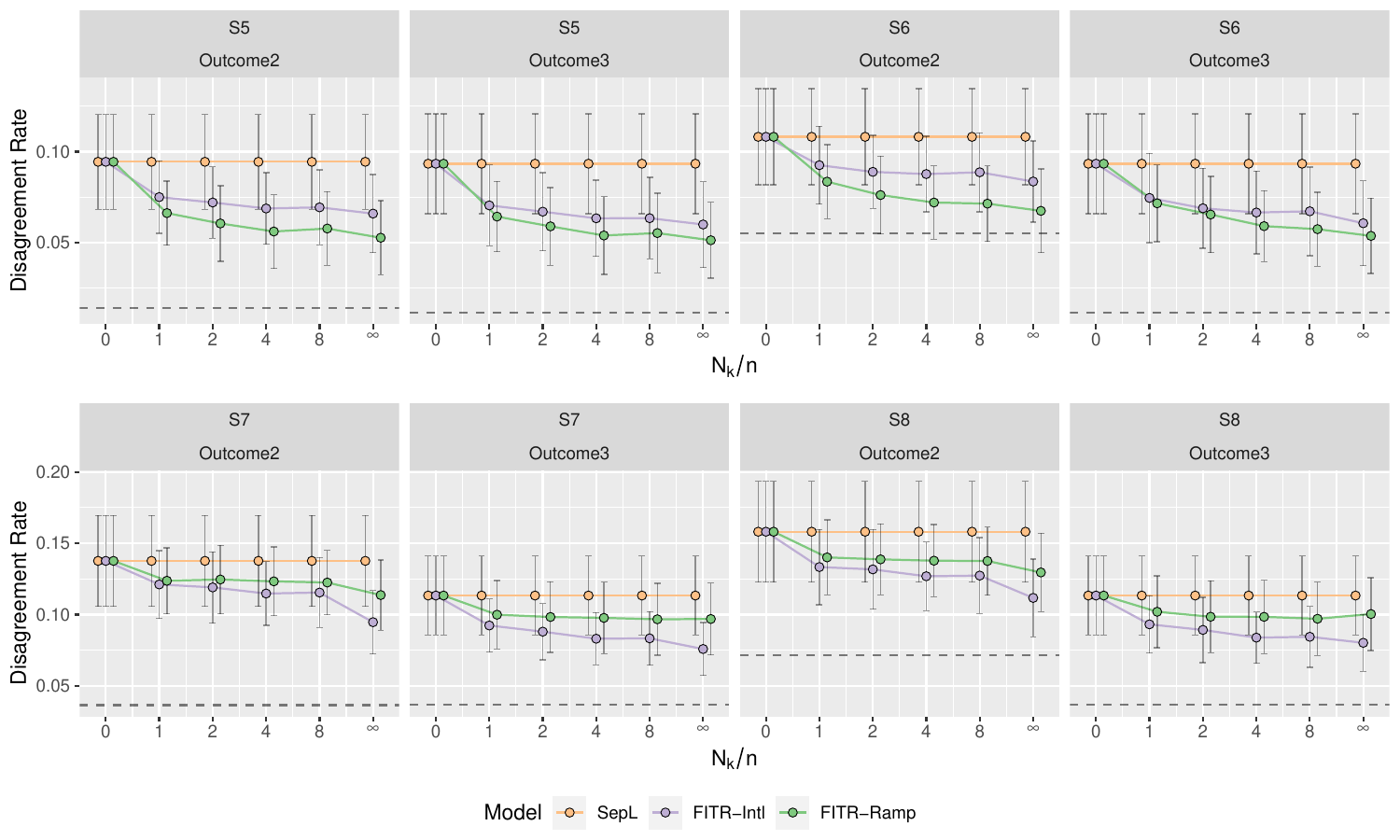}
    \caption{The mean and standard deviation of disagreement rate between FITR $\widehat{f}_{1n}$ learned using SepL, FITR-Ramp or FITR-IntL, and the true secondary outcome ITRs $f_2^*, f_3^*$ when $K = 3$ and $n = 200$. The subtitle of each subfigure refers to the secondary outcome for which we are estimating the disagreement rate. For example, ``Outcome3'' refers to the disagreement rate $\bbP(\widehat{f}_{1n} f_3^* < 0)$. The dashed line represents the true disagreement rate $\bbP(f_1^* f_2^* < 0)$ or $\bbP(f_1^* f_3^* < 0)$.}
    \label{fig:K3.disagree.n200}
\end{figure}

\begin{figure}[p]
    \centering
    \begin{subfigure}{0.45\textwidth}
        \centering
        \includegraphics[width=\textwidth]{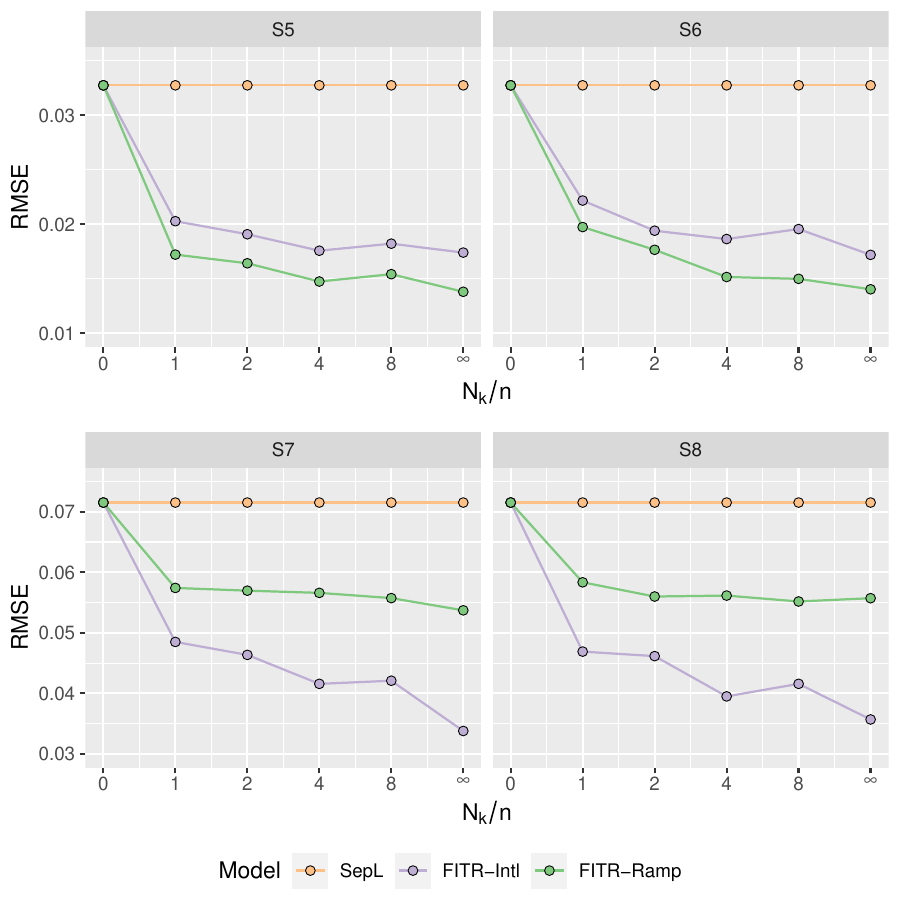}
        \caption{RMSE}
    \end{subfigure}
    \hfill
    \begin{subfigure}{0.45\textwidth}
        \centering
        \includegraphics[width=\textwidth]{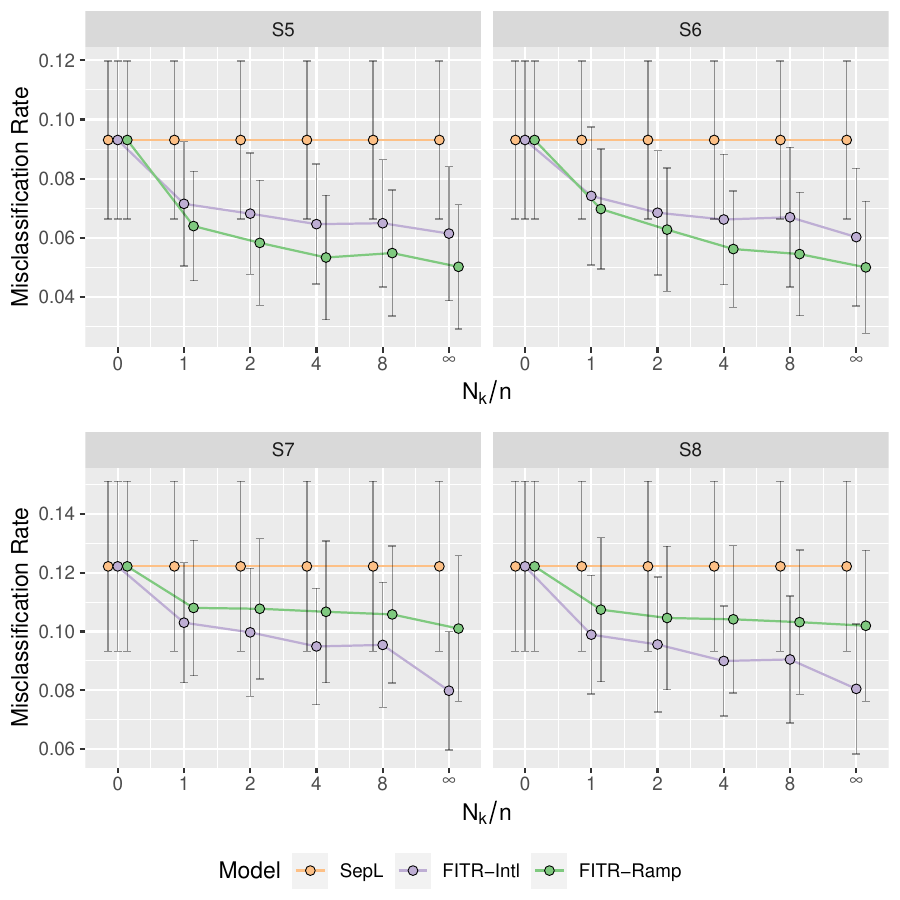}
        \caption{Accuracy}
    \end{subfigure}
    \caption{(a) The RMSE of value functions and (b) the mean and standard deviation of accuracy when FITR is learned using SepL, FITR-Ramp or FITR-IntL when $K = 3$ and $n = 200$.}
    \label{fig:K3.rmse.accuracy.n200}
\end{figure}

\begin{figure}[p]
    \centering
    \includegraphics[width=0.8\textwidth]{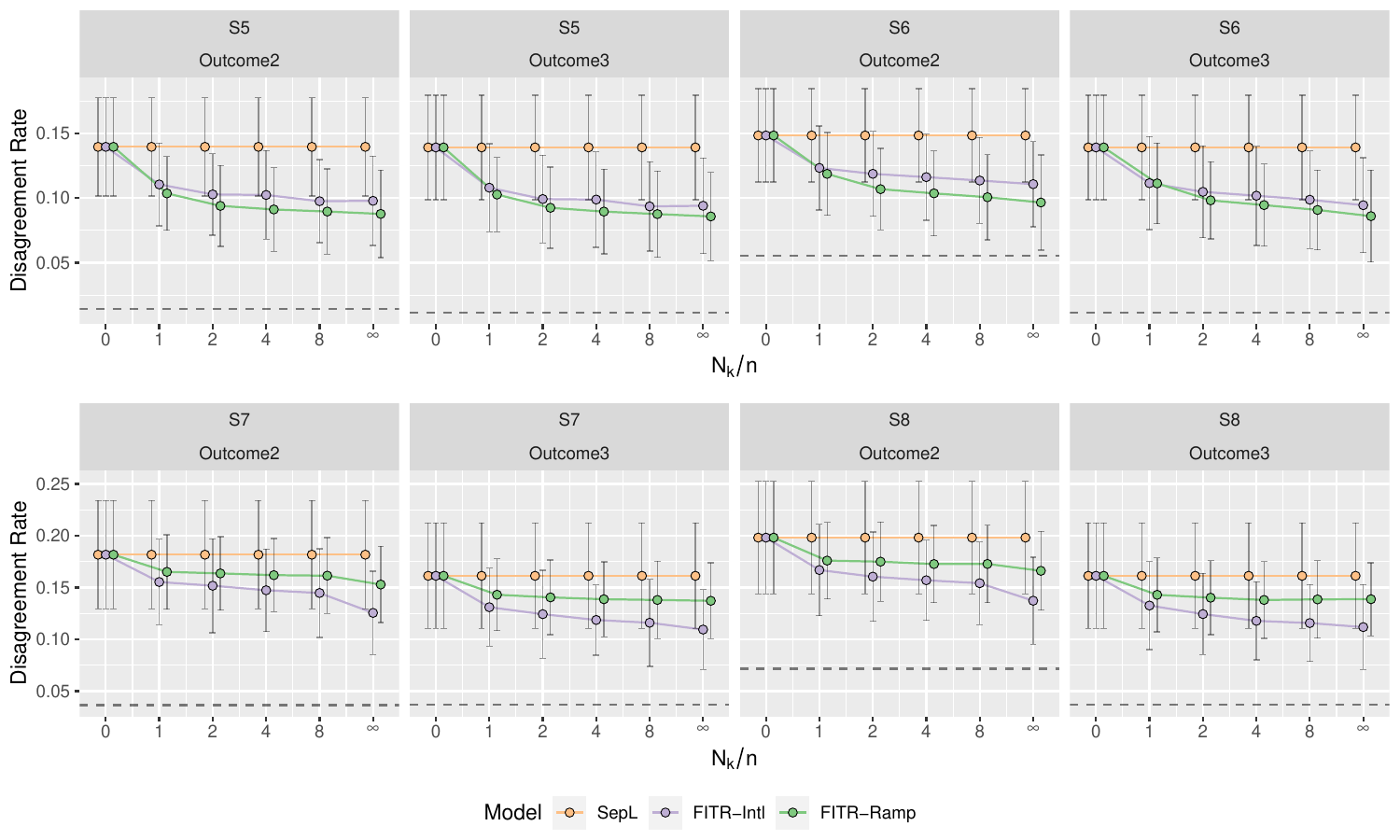}
    \caption{The mean and standard deviation of disagreement rate between FITR $\widehat{f}_{1n}$ learned using SepL, FITR-Ramp or FITR-IntL, and the true secondary outcome ITRs $f_2^*, f_3^*$ when $K = 3$ and $n = 100$. The subtitle of each subfigure refers to the secondary outcome for which we are estimating the disagreement rate. For example, ``Outcome3'' refers to the disagreement rate $\bbP(\widehat{f}_{1n} f_3^* < 0)$. The dashed line represents the true disagreement rate $\bbP(f_1^* f_2^* < 0)$ or $\bbP(f_1^* f_3^* < 0)$.}
    \label{fig:K3.disagree.n100}
\end{figure}

\begin{figure}[p]
    \centering
    \begin{subfigure}{0.45\textwidth}
        \centering
        \includegraphics[width=\textwidth]{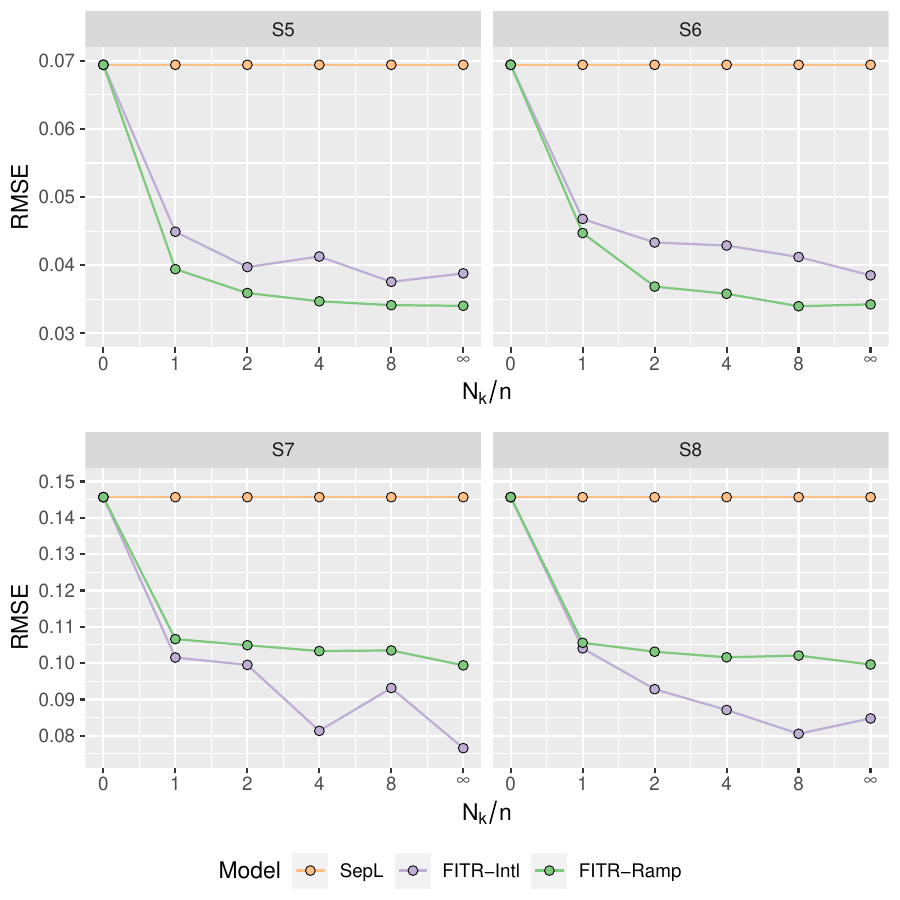}
        \caption{RMSE}
    \end{subfigure}
    \hfill
    \begin{subfigure}{0.45\textwidth}
        \centering
        \includegraphics[width=\textwidth]{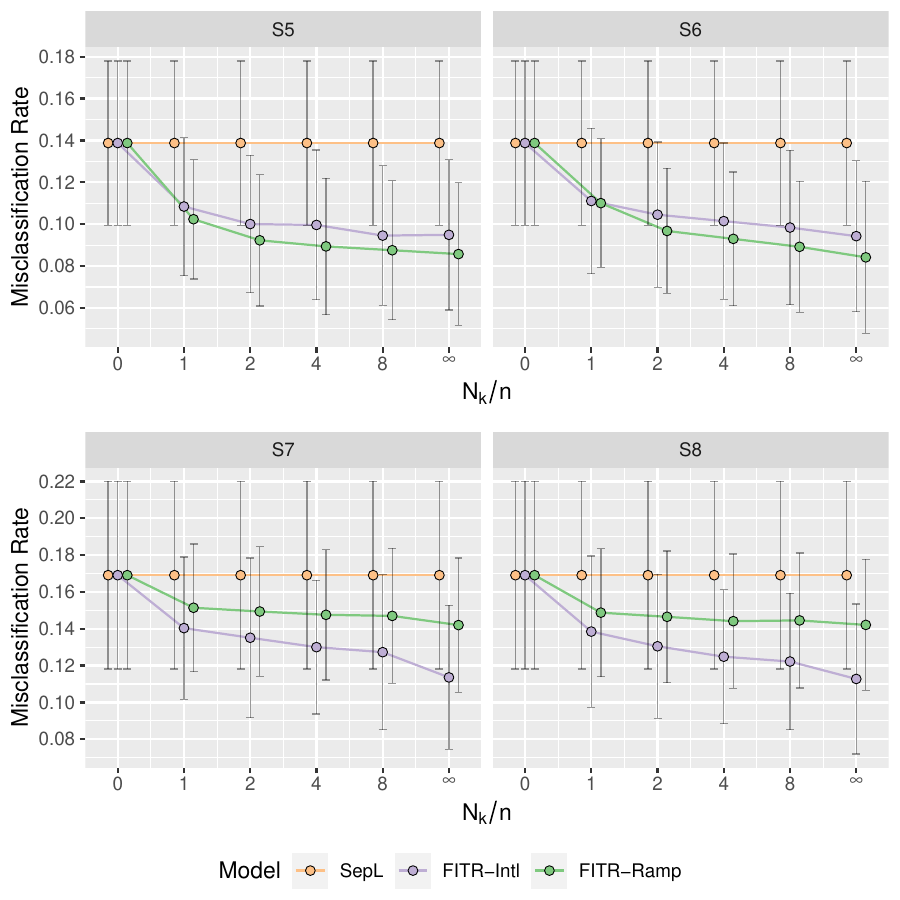}
        \caption{Accuracy}
    \end{subfigure}
    \caption{(a) The RMSE of value functions and (b) the mean and standard deviation of accuracy when FITR is learned using SepL, FITR-Ramp or FITR-IntL when $K = 3$ and $n = 100$.}
    \label{fig:K3.rmse.accuracy.n100}
\end{figure}

Figures~\ref{fig:K3.disagree.n200} and ~\ref{fig:K3.rmse.accuracy.n200} contains the simulation results when $K = 3$ and $n = 200$, and Figures~\ref{fig:K3.disagree.n100} and ~\ref{fig:K3.rmse.accuracy.n100} contains the simulation results when $K = 3$ and $n = 100$.
When the number of outcomes $K$ increases, the disagreement rate, the RMSE and misclassification rate all decreases.
Although the discrepancy between $R_1$ and $R_2$ is larger in S6 than in S5, the RMSE and misclassification rate of $R_1$ are almost the same as in S5.
Remember that RMSE and misclassification rate are larger in S2 than in S1, although $R_1$ and $R_2$ are exactly the same as in S5 and S6 correspondingly.
This reflects the advantage of using the third outcome, in spite of the fact that $R_2$ and $R_3$ differ from $R_1$ in opposite directions.

\subsection{Additional Results for Sensitivity Analysis in Section~\ref{sec:simulation}} \label{suppsec:sub.sensitivity}

This section contains the details of the sensitivity analysis in Section~\ref{sec:simulation}.
In Table~\ref{tbl:sensitivity}, we show different values of $\rho$ and their corresponding agreement rates between $f_1^*$ and $f_2^*$ when $n = 200$ and $N_2 = 4n$.

\begin{table}[!ht]
    \small
    \centering
    \caption{The change of RMSE and accuracy when the similarity between outcomes is changed.}
    \setlength{\tabcolsep}{0pt}
    \begin{tabular*}{\linewidth}{@{\extracolsep{\fill}} ccccccc}
        \toprule
        $\rho$ & $\bbP(f_1^* f_2^* > 0)$ & Model & $\bbP(\widehat{f}_{1n} f_2^* > 0)$ & RMSE &$\frac{\text{RMSE}}{\text{RMSE}_{SepL}}$ & Misclassification Rate \\
        \midrule
        \multirow{3}{*}{1} & \multirow{3}{*}{100\%}
          & SepL      & 0.905 (0.026) & 0.034 & 1.000 & 0.095 (0.026) \\ 
        & & FITR-IntL & 0.925 (0.024) & 0.023 & 0.682 & 0.075 (0.024) \\ 
        & & FITR-Ramp & 0.936 (0.022) & 0.019 & 0.569 & 0.064 (0.022) \\ 
        \midrule
        \multirow{3}{*}{0.9} & \multirow{3}{*}{98.57\%}
          & SepL      & 0.904 (0.025) & 0.034 & 1.000 & 0.095 (0.026) \\ 
        & & FITR-IntL & 0.922 (0.022) & 0.023 & 0.679 & 0.076 (0.024) \\ 
        & & FITR-Ramp & 0.935 (0.022) & 0.019 & 0.557 & 0.064 (0.022) \\ 
        \midrule
        \multirow{3}{*}{0.8} & \multirow{3}{*}{96.81\%}
          & SepL      & 0.900 (0.025) & 0.034 & 1.000 & 0.095 (0.026) \\ 
        & & FITR-IntL & 0.916 (0.023) & 0.024 & 0.727 & 0.078 (0.025) \\ 
        & & FITR-Ramp & 0.928 (0.023) & 0.021 & 0.616 & 0.070 (0.023) \\ 
        \midrule
        \multirow{3}{*}{0.7} & \multirow{3}{*}{94.48\%}
          & SepL      & 0.891 (0.025) & 0.034 & 1.000 & 0.095 (0.026) \\ 
        & & FITR-IntL & 0.909 (0.024) & 0.026 & 0.770 & 0.081 (0.026) \\ 
        & & FITR-Ramp & 0.920 (0.026) & 0.024 & 0.698 & 0.076 (0.023) \\ 
        \midrule
        \multirow{3}{*}{0.6} & \multirow{3}{*}{91.42\%}
          & SepL      & 0.874 (0.027) & 0.034 & 1.000 & 0.095 (0.026) \\ 
        & & FITR-IntL & 0.895 (0.027) & 0.028 & 0.834 & 0.086 (0.026) \\ 
        & & FITR-Ramp & 0.908 (0.030) & 0.027 & 0.811 & 0.085 (0.023) \\ 
        \midrule
        \multirow{3}{*}{0.5} & \multirow{3}{*}{87.56\%}
          & SepL      & 0.847 (0.029) & 0.034 & 1.000 & 0.095 (0.026) \\ 
        & & FITR-IntL & 0.873 (0.032) & 0.033 & 0.970 & 0.094 (0.028) \\ 
        & & FITR-Ramp & 0.888 (0.038) & 0.033 & 0.976 & 0.096 (0.025) \\ 
        \midrule
        \multirow{3}{*}{0.4} & \multirow{3}{*}{83.16\%}
          & SepL      & 0.810 (0.031) & 0.034 & 1.000 & 0.095 (0.026) \\ 
        & & FITR-IntL & 0.843 (0.040) & 0.037 & 1.086 & 0.101 (0.028) \\ 
        & & FITR-Ramp & 0.858 (0.046) & 0.039 & 1.169 & 0.106 (0.029) \\ 
        \bottomrule
    \end{tabular*}
    \label{tbl:sensitivity}
\end{table}

\subsection{Additional Results for Real Data Analysis in Section~\ref{sec:realdata}} \label{suppsec:sub.realdata}

Two covariates were shown to be informative for tailoring treatments in a prior study \citep{chen2021learning}.
The first is the NEO-Five Factor Inventory score, where the NEO Personality Inventory is a 240-item measurement designed to assess personality in the domains of neuroticism, extraversion, openness, and so on; we focus on the neuroticism domain. 
The second informative measure is the Flanker Interference Accuracy score, where a higher value indicates reduced cognitive control. 
Five additional baseline variables are used, including sex, age, education years, Edinburgh Handedness Inventory (EHI) score, and the QIDS score at the beginning of the study. 

We list the coefficients of the ITR $\widehat{f}_{\text{QIDS}}$ of the primary outcome estimated using SepL, FITR-IntL and FITR-Ramp in Table~\ref{tbl:EMBARC.itr}, along with the secondary outcome ITRs $\widetilde{f}_{\text{CGI}}$ and $\widetilde{f}_{\text{SAS}}$ estimated using SepL.
They are learned with the complete dataset and the linear kernel.
We can conclude that the coefficients fitted by different methods are generally close, which suggests that the fusion penalty will not dramatically change an ITR compared to SepL. 
The coefficients are more similar for QIDS-change and CGI, which is expected since SAS measures the impacts of depression on social functioning. 
The Flanker variable has a different sign in the ITR of SAS than in the other two outcomes.

\begin{table}[!ht]
    \centering
    \caption{Coefficients of the estimated ITRs of the three outcomes when linear kernel is used.}
    \begin{tabular*}{\linewidth}{@{\extracolsep{\fill}} cccccccccc}
        \toprule
        Outcome & Model & intercept & sex & age & education & EHI & QIDS & NEO & Flanker \\
        \midrule
        \multirow{3}{*}{\makecell{QIDS-\\change}} 
        & SepL      & 0.130 & 0.105 & 0.231 & -0.126 & 0.066 & 0.031 & 0.251 & -0.345 \\ 
        & FITR-IntL & 0.137 & 0.098 & 0.126 & -0.020 & 0.032 & 0.065 & 0.087 & -0.174 \\ 
        & FITR-Ramp & 0.063 & 0.038 & 0.078 & -0.036 & 0.025 & 0.027 & 0.071 & -0.108 \\ 
        \midrule
        \multirow{1}{*}{CGI} 
        & SepL      & 0.358 & 0.019 & 0.275 & -0.043 & 0.086 & 0.014 & 0.155 & -0.281 \\ 
        \midrule
        \multirow{1}{*}{SAS} 
        & SepL      & 0.207 & 0.014 & 0.147 & -0.102 & -0.051 & 0.202 & 0.157 & 0.029 \\ 
        \bottomrule
    \end{tabular*}
    \label{tbl:EMBARC.itr}
\end{table}

\section{ADDITIONAL ASSUMPTIONS, RESULTS AND PROOFS FOR SECTION~\ref{sec:theory}} \label{suppsec:theo}

In this section, we provide additional assumptions, theoretical results and proofs.
The proofs are extended from outcome weighted learning with a hinge loss to our problem with logistic loss
and nonconvex ramp loss.
Besides, we also used a new change of measure technique in Section~\ref{suppsec:sub.proof.value}, and derived a new result in Corollary~\ref{thm:convergence.rate.accuracy} under a relatively weak Assumption~\ref{asp:nonzero.rewardsum}. 



\subsection{Additional Assumptions and Theoretical Results} \label{suppsec:sub.theory}

We present the details about the ignorability, consistency, and positivity assumptions, which are common in causal inference to identify the treatment effect.

\begin{asp}[Ignorability] \label{asp:ignorability}
    The treatment $A$ is independent of the potential outcomes $R_1^* (a)$ given covariates $\bX$.
\end{asp}

\begin{asp}[Consistency] \label{asp:consistency}
    The observed outcome $R_1$ under a treatment $A = a$ equals the potential outcome $R_1 (a)$ for all $a \in \cA$.
\end{asp}

\begin{asp}[Positivity] \label{asp:positivity}
    There exists $p_0 > 0$ such that $\pi(a; \bx)\equiv P(A=a|\bX=\bx) > p_0$ for all $a \in \cA$ and all $\bx \in \cX$.
\end{asp}

To bound the risk with 0-1 loss using the risk with convex surrogate loss, we need another common assumption in classification literature, the Tsybakov's noise assumption, which assumes the noise conditions around the decision boundary \citep{bartlett2006convexity}.

\begin{asp} \label{asp:Tsybakov.noise}
    Assume that the distribution of $\bX$ satisfies the following Tsybakov's noise assumption:
    there exists a constant $C > 0$ such that for all sufficiently small $t > 0$ we have
    \[ \bbP ( \brce{\bX \in \cX: \abs{2 \eta(\bX) - 1} \le t}) \le C  t^{\beta} \] for some $\beta>0$.
    Let $\alpha = \beta / (1 + \beta)$ so that $\alpha \in (0, 1]$.
\end{asp}

For Assumption~\ref{asp:Tsybakov.noise}, it has been shown that the boundary assumption of $\eta(\bx)$ regarding $\beta$ is equivalent to the misclassification assumption of $f_1$ regarding $\alpha$ \citep{bartlett2006convexity}.

Assumption~\ref{asp:nonzero.rewardsum} is necessary when bounding the decision accuracy.
It is a weak assumption that only requires one of the treatments to yield a positive mean reward, which is easily achievable by shifting the original rewards with a positive constant

\begin{asp} \label{asp:nonzero.rewardsum}
    Suppose the conditional expectation of rewards satisfies $\sum_{a \in \cA} r_1^{(a)} (\bx) \ge c_r$ for some constant $c_r > 0$ for all $\bx \in \cX$.
\end{asp}

Now we can present the convergence rates of the value function $\cV_1 (\widehat{f}_{1n})$ and the misclassification rate $\bbP (\widehat{f}_{1n} f_1^* < 0)$ when $K = 2$. 

\begin{thm} \label{thm:convergence.rate.value}
    Under Assumptions~\ref{asp:ignorability}-\ref{asp:Tsybakov.noise} and~\ref{asp:bounded.reward}-\ref{asp:conv.secondary}, 
    when $K = 2$,
    the value function of the estimated FITR satisfies
    \begin{equation} \label{equ:convergence.rate.value}
        \cV_1 (f_1^*) - \cV_1 (\widehat{f}_{1n}) \lesssim (\delta_{1n}(\tau) + \mu_{1n})^{\frac{1}{2 - \alpha}}, 
    \end{equation}
    with probability at least $1 - 2 e^{\tau}$.
\end{thm}

\begin{coro} \label{thm:convergence.rate.accuracy}
    Under Assumptions~\ref{asp:ignorability}-\ref{asp:nonzero.rewardsum} and~\ref{asp:bounded.reward}-\ref{asp:conv.secondary}, 
    the misclassification rate satisfies
    \begin{equation} \label{equ:convergence.rate.accuracy}
        \bbP (\widehat{f}_{1n} f_1^* < 0) \lesssim (\delta_{1n}(\tau) + \mu_{1n})^{\frac{\alpha}{2 - \alpha}}
    \end{equation}
    with probability at least $1 - 2 e^{\tau}$.
\end{coro}

Notice that the convergence rate $\widetilde{\delta}_{2 N_2}(\tau)$ of $\widetilde{f}_2$ does not appear in (\ref{equ:convergence.rate.value}), since the term containing $\widetilde{\delta}_{2 N_2}(\tau)$ is not dominant in the proof due to the assumption that $\widetilde{\delta}_{2 N_2}(\tau) = o(1)$.
By comparing $\delta_{1n}(\tau) + \mu_{1n}$ and $\delta_{1n}^{(0)}(\tau)$, it can be concluded that FITR has a slower convergence rate for the value function than SepL. This is due to the fact that the fusion penalty introduces bias for maximizing the primary outcome.

The results in Theorem~\ref{thm:convergence.rate.value} and Corollary~\ref{thm:convergence.rate.accuracy} can also be generalized to $K \ge 2$.

\begin{coro} \label{thm:convergence.rate.value.accuracy.K}
    Under Assumptions~\ref{asp:ignorability}-\ref{asp:Tsybakov.noise} and~\ref{asp:bounded.reward}-\ref{asp:conv.secondary}, 
    when $K \ge 2$,
    the value function of the estimated FITR $\widehat{f}_{1n}$ satisfies
    \begin{equation} \label{equ:convergence.rate.value.K}
        \cV_1 (f_1^*) - \cV_1 (\widehat{f}_{1n}) \lesssim (\delta_{1n}(\tau) + \mu_{1n})^{\frac{1}{2 - \alpha}},
    \end{equation}
    with probability at least $1 - K e^{\tau}$.

    Under Assumptions~\ref{asp:ignorability}-\ref{asp:nonzero.rewardsum} and~\ref{asp:bounded.reward}-\ref{asp:conv.secondary}, 
    the misclassification rate satisfies
    \begin{equation} \label{equ:convergence.rate.accuracy.K}
        \bbP (\widehat{f}_{1n} f_1^* < 0) \lesssim (\delta_{1n}(\tau) + \mu_{1n})^{\frac{\alpha}{2 - \alpha}}
    \end{equation}
    with probability at least $1 - K e^{\tau}$.
\end{coro}

\begin{rmk}
    If $\widetilde{f}_k$ is the sign of the decision function learned by SepL with sample size $N_k$,
    we can directly use Corollary~\ref{thm:convergence.rate.accuracy} with $\mu_{k N_k} = 0$ to find that $\widetilde{\delta}_{k N_k}(\tau)$ in Assumption~\ref{asp:conv.secondary} can be expressed as 
    $\widetilde{\delta}_{k N_k}(\tau) = (\delta_{k N_k}^{(0)}(\tau))^{\frac{\alpha}{2 - \alpha}}$,
    where 
    \[ \delta_{k N_k}^{(0)}(\tau) :=
    \lambda_{k N_k}^{-\frac{1}{2}}  N_k^{-\frac{1}{2}} \brck{\sqrt{\tau} + \sigma_{k N_k}^{(1 - \nu/2) (1 + \delta) d}}
    + \lambda_{k N_k} \sigma_{k N_k}^d + (2d)^{q d / 2} \sigma_{k N_k}^{-q d}. \]
\end{rmk}

\begin{rmk}
    With a stronger assumption $\absn{r_1^{(1)} (\bx) - r_1^{(-1)} (\bx)} > 0$ for all $\bx$, which assumes that the treatment effect is nonzero for all patients, $\bbP (\widehat{f}_{1n} f_1^* < 0)$ can be directly bounded by a multiple of $\cV_1 (f_1^*) - \cV_1 (\widehat{f}_{1n})$ in Theorem~\ref{thm:convergence.rate.value}.
    Corollary~\ref{thm:convergence.rate.accuracy} does not rely on the assumption and only assumes that $r_1^{(a)} (\bx)$ cannot be zero at the same time for all $a \in \cA$.
    Under the weaker assumption, the convergence rate of $\bbP (\widehat{f}_{1n} f_1^* < 0)$ is slower with a scaling constant $\alpha \in (0,1]$.
\end{rmk}

\subsection{Analysis Details for Section~\ref{sec:theory}} \label{suppsec:sub.details.optimal.rates}

The parameters $\lambda_{1n}, \mu_{1n}, \kappa_{1n}$ should balance the estimation error and approximation error to achieve the minimal upper bound $\delta_{1n}(\tau)$.
The minimal approximation error is
$\gamma_{1n}^{\frac{1}{1+q}} \lambda_{1n}^{\frac{q}{1+q}}$
when
$\sigma_{1n} = \brce{\gamma_{1n} \lambda_{1n}^{-1}}^{\frac{1}{(1+q) d}}$.
Then the best $\lambda_{1n}$ that balances the estimation error and the approximation error is 
$\lambda_{1n} = \gamma_{1n}^{\frac{2 (\Delta - 1) + 2 (1 + q) \nu}{3q + 2\Delta + 1}} n^{- \frac{1 + q}{3q + 2\Delta + 1}}$.
When $\mu_{1n} = 0$, the convergence rate of the risk is
\begin{equation*}
    \begin{split}
        \delta_{1n}^{(0)}(\tau) :=
        \lambda_{1n}^{-\frac{1}{2}} n^{-\frac{1}{2}} \brck{\sqrt{\tau} + \sigma_{1n}^{(1 - \nu/2) (1 + \delta) d}} 
        + \lambda_{1n} \sigma_{1n}^d + (2d)^{q d / 2} \sigma_{1n}^{-q d}.
    \end{split}
\end{equation*}
The minimum approximation error $\lambda_{1n}^{\frac{q}{1+q}}$ is obtained when 
$\sigma_{1n} = \lambda_{1n}^{-\frac{1}{(1+q) d}}$.
Then the best $\lambda_{1n}$ is 
$n^{- \frac{1+q}{3q + 2\Delta + 1}}$.

To bound the agreement rate without the fusion penalty, first note that
$$
    \bbP (\widehat{f}_{1n} f_2^* < 0) 
    = \bbP (\widehat{f}_{1n} f_1^* < 0, f_1^* f_2^* > 0) + \bbP (\widehat{f}_{1n} f_1^* > 0, f_1^* f_2^* < 0)
    \le \bbP (\widehat{f}_{1n} f_1^* < 0) + \bbP (f_1^* f_2^* < 0),
$$
which allows us to obtain the bound with the help of the misclassification rate.
For SepL with $\mu_{1n} = 0$, we have
$\bbP (\widehat{f}_{1n} f_1^* < 0) \lesssim n^{- \frac{\alpha}{2 - \alpha} \frac{q}{3q + 2\Delta + 1}}$
with probability at least $1 - 2 e^{\tau}$ 
according to (\ref{equ:min.delta.n0}) and Corollary \ref{thm:convergence.rate.accuracy}.
Therefore, we can conclude that for SepL
\begin{equation}
    \begin{split}
        \bbP (f_1^* f_2^* > 0) - \bbP (\widehat{f}_{1n} f_2^* > 0)
        \le \bbP (\widehat{f}_{1n} f_1^* < 0) 
        \lesssim n^{- \frac{\alpha}{2 - \alpha} \frac{q}{3q + 2\Delta + 1}}.
    \end{split}
\end{equation}

To compare the agreement rate with or without the fusion penalty, we break down equation (\ref{equ:compare.SepL.FITR}) as follows.
When $\mu_{1n} \gtrsim \kappa_{1n}$, we have $\gamma_{1n} \simeq \mu_{1n} / \kappa_{1n}$.
If $\alpha = 1/2, \nu = 3/2, \delta = 1, \Delta = 1/2$, the first condition in (\ref{equ:compare.SepL.FITR}) is satisfied with $\kappa_{1n} \gtrsim n^{- 2q / (9q+6)}, \mu_{1n} \gtrsim \kappa_{1n}$, and $N_2 \gtrsim n^3 \kappa_{1n}^{(9q+6) / q} \gtrsim n$.
On the other hand, when $\mu_{1n} \lesssim \kappa_{1n}$, we have $\gamma_{1n} \simeq 1$.
For the same $\alpha, \nu, \delta, \Delta$, the second condition in (\ref{equ:compare.SepL.FITR}) is satisfied with $\mu_{1n} \gtrsim n^{- 2q / (9q+6)}, \kappa_{1n} \gtrsim \mu_{1n}$, and $N_2 \gtrsim n^3 \mu_{1n}^{(9q+6) / q} \gtrsim n$.

\subsection{Proof of Lemma~\ref{lem:surrogate.risk}} \label{suppsec:sub.proof.surrogate.risk}

\begin{proof}
    First note that
    \begin{align}
        & \cR (\widehat{f}_{1n}) - \cR (f_1^*) 
        \le \cR (\widehat{f}_{1n}) - \cR (f_1^*) + \lambda_{1n} \normn{\widehat{f}_{1n}}_{\cH}^2 \\
        \le & \brce{ \brck{ \cR (\widehat{f}_{1n}) + \lambda_{1n} \normn{\widehat{f}_{1n}}_{\cH}^2 } - \inf_{f_1 \in \cH} \brck{ \cR (f_1) + \lambda_{1n} \normn{f_1}_{\cH}^2 }} \label{equ:estimation.error} \\ 
        & + \brce{ \inf_{f_1 \in \cH} \brck{ \cR (f_1) + \lambda_{1n} \normn{f_1}_{\cH}^2 } - \cR (f_1^*) }. \label{equ:approx.error}
    \end{align}
    Define $f_1^{\dagger}$ as the minimizer of $\cR (f_1) + \lambda_{1n} \normn{f_1}_{\cH}^2$ in $\cH$.
    We will bound the two terms on the right-hand side separately.

    To bound (\ref{equ:approx.error}), we follow the construction in the proof of Theorem of 2.7 in \cite{steinwart2007fast}.
    When $\cX$ is the closed unit ball, on $\acute{\cX} := 3 \cX$ define 
    \begin{equation*}
    \acute{\eta} (\bx) =
    \begin{cases}
        \eta (\bx), & \text{if } \norm{\bx}_2 \le 1, \\
        \eta (\bx / \norm{\bx}_2), & \text{otherwise}.
    \end{cases}
    \end{equation*}
    Besides, let $\acute{\cX}_{-1} := \{ x \in \acute{\cX}: \acute{\eta} (\bx) < \frac{1}{2} \}$ and $\acute{\cX}_{1} := \{ x \in \acute{\cX}: \acute{\eta} (\bx) > \frac{1}{2} \}$.
    Fix a measurable $\acute{f}_1 : \acute{\cX} \mapsto [-1, 1]$ that satisfies $\acute{f}_1 = 1$ on $\acute{\cX}_{1}$, $\acute{f}_1 = -1$ on $\acute{\cX}_{-1}$ and $\acute{f}_1 = 0$ otherwise.
    The linear operator $V_{\sigma_{1n}}: L_2 (\bbR^d) \mapsto \cH_{\sigma_{1n}} (\bbR^d)$ defined by
    \[ V_{\sigma_{1n}} g(\bx) = \frac{(2 \sigma_{1n})^{d/2}}{\pi^{d/4}} \int_{\bbR^d} e^{- 2 \sigma_{1n}^2 \norm{\bx - \by}_2^2} g(\by) d \by, \qquad g \in L_2 (\bbR^d), \bx \in \bbR^d, \]
    is an isometric isomorphism \citep{steinwart2006explicit}.
    Consequently, we have
    \begin{equation*}
    \begin{split}
        & \inf_{f_1 \in \cH} \brck{ \cR (f_1) + \lambda_{1n} \normn{f_1}_{\cH}^2 } - \cR (f_1^*) \\
        \le & \inf_{g \in L_2(\bbR^d)} \brck{ \bbE (\ell_1 \circ V_{\sigma_{1n}} g - \ell_1 \circ f_1^*) + \bbE (\ell_2 \circ V_{\sigma_{1n}} g - \ell_2 \circ f_1^*) + \lambda_{1n} \normn{g}_{L_2 (\bbR^d)}^2 }.
    \end{split}
    \end{equation*}
    Now take a specific $g := \big(\frac{\sigma_{1n}^2}{\pi}\big)^{d/4} \acute{f}_1$, and we obtain
    \begin{equation} \label{equ:approx.g.L2norm}
        \normn{g}_{L_2 (\bbR^d)} \le \prth{\frac{81 \sigma_{1n}^2}{\pi}}^{d/4} \theta (d),
    \end{equation}
    where $\theta (d)$ denotes the volume of $\cX$.
    Since $\-1 \le \acute{f}_1 \le 1$, it can be easily seen that $\-1 \le V_{\sigma_{1n}} g \le 1$.
    Note that $\abs{\phi^{\prime} (t)} = \abs{- \frac{e^{-t}}{1 + e^{-t}}} \le 1$, so $\ell_1$ is Lipschitz continuous with respect to $f_1$ with Lipschitz constant $r / p_0$.
    It has been shown in \cite{steinwart2007fast} that 
    \[ \abs{V_{\sigma_{1n}} g (\bx) - f_1^* (\bx)} \le 8 e^{- \sigma_{1n}^2 \omega_{\bx}^2 / 2d}. \]
    Therefore, Assumption~\ref{asp:geometric.noise} for $t = 2d / \sigma_{1n}^2$ yield
    \begin{equation} \label{equ:approx.logistic}
        \bbE (\ell_1 \circ V_{\sigma_{1n}} g - \ell_1 \circ f_1^*) \lesssim \bbE \abs{V_{\sigma_{1n}} g - f_1^*} \lesssim \bbE e^{- \sigma_{1n}^2 \omega_{\bx}^2 / 2d} \lesssim (2d)^{q d / 2} \sigma_{1n}^{-q d}.
    \end{equation}
    Since $\ell_2$ is Lipschitz continuous with respect to $f_1$ with Lipschitz constant $\frac{\mu_{1n} \Omega_{12}}{\kappa_{1n}} \normn{\widetilde{f}_2}_{\infty}$,
    \begin{equation} \label{equ:approx.ramp}
        \bbE (\ell_2 \circ V_{\sigma_{1n}} g - \ell_2 \circ f_1^*) \lesssim \frac{\mu_{1n}}{\kappa_{1n}} \bbE \abs{V_{\sigma_{1n}} g - f_1^*} \lesssim \frac{\mu_{1n}}{\kappa_{1n}} (2d)^{q d / 2} \sigma_{1n}^{-q d}
    \end{equation}
    when $\normn{\widetilde{f}_2}_{\infty} = 1$.
    Combining (\ref{equ:approx.g.L2norm}), (\ref{equ:approx.logistic}) and (\ref{equ:approx.ramp}), we can bound the approximation error (\ref{equ:approx.error}) as 
    \begin{equation} \label{equ:approx.error.result}
        \inf_{f_1 \in \cH} \brck{ \cR (f_1) + \lambda_{1n} \normn{f_1}_{\cH}^2 } - \cR (f_1^*)
    \lesssim \lambda_{1n} \sigma_{1n}^d + \gamma_{1n} (2d)^{q d / 2} \sigma_{1n}^{-q d},
    \end{equation}
    where $\gamma_{1n} := 1 + \mu_{1n} / \kappa_{1n}$.

    To bound (\ref{equ:estimation.error}), we will use Talagrand's inequality quoted as follows \citep[Theorem 5.3]{steinwart2007fast}.

    \begin{thm} \label{thm.Talagrand}
        Let $\cH$ be a set of bounded measurable functions from $\bZ$ to $\bbR$, which is separable with respect to $\norm{\cdot}_{\infty}$ and satisfies $\bbE h = 0$ for all $h \in \cH$. 
    Furthermore, let $B > 0$ and $b \ge 0$ be constants with $\norm{h}_{\infty} \le B$ and $\bbE h^2 \le b$ for all $h \in \cH$. 
    Then for all $\tau \ge 1$ and all $n \ge 1$ we have
    \[ \bbP \prth{\sup_{h \in \cH} \bbP_n h > 3 \bbE \sup_{h \in \cH} \bbP_n h + \sqrt{\frac{2 \tau b}{n}} + \frac{B \tau}{n}} \le e^{- \tau}. \]
    \end{thm}

    We first obtain a bound for $\normn{\widehat{f}_{1n}}_{\cH}^2$.
    Since $\bbP_n (\ell_1 \circ \widehat{f}_{1n} + \ell_2 \circ \widehat{f}_{1n}) + \lambda_{1n} \normn{\widehat{f}_{1n}}_{\cH}^2 \le \bbP_n (\ell_1 \circ f_1 + \ell_2 \circ f_1) + \lambda_{1n} \normn{f_1}_{\cH}^2$ for any $f_1 \in \cH$, when taking $f_1 = 0$ we have 
    \[ \lambda_{1n} \normn{\widehat{f}_{1n}}_{\cH}^2 \le \bbP_n (\ell_1 \circ f_1 + \ell_2 \circ f_1) \le \frac{r}{p_0} + \mu_{1n} \Omega_{12}. \]
    Since $r / p_0 + \mu_{1n} \Omega_{12} \simeq M$ with $M := r / p_0$, $\normn{\widehat{f}_{1n}}_{\cH}$ is bounded by $\sqrt{M / \lambda_{1n}}$.
    To this end, it suffices to consider the ball of radius $\sqrt{M / \lambda_{1n}}$.
    Therefore, the function class that we consider here is 
    \[ \cG := \brce{\ell_1 \circ f_1 + \ell_2 \circ f_1 + \lambda_{1n} \normn{f_1}_{\cH}^2 - \brck{\ell_1 \circ f_1^{\dagger} + \ell_2 \circ f_1^{\dagger} + \lambda_{1n} \normn{f_1^{\dagger}}_{\cH}^2}: f \in B_{\cH} (\sqrt{M / \lambda_{1n}})}, \]
    where $B_{\cH} (r)$ is the ball in $\cH$ of radius $r$.
    Since $\ell_1$ is Lipschitz continuous with respect to $f_1$
    and $\norm{f}_{\infty} \le \norm{f}_{\cH}$ for any $g \in \cG$, 
    \begin{equation*}
    \begin{split}
        \abs{g} 
        \le & \abs{\ell_1 \circ f_1 - \ell_1 \circ f_1^{\dagger}} + \abs{\ell_2 \circ f_1 - \ell_2 \circ f_1^{\dagger}} + \lambda_{1n} \abs{\normn{f_1}_{\cH}^2 - \normn{f_1^{\dagger}}_{\cH}^2} \\
        \le & M \abs{f_1 - f_1^{\dagger}} + \mu_{1n} \Omega_{12} + M \\
        \le & 2 M \sqrt{M / \lambda_{1n}} + \mu_{1n} \Omega_{12} + M.
    \end{split}
    \end{equation*}
    Hence, with $B := 2 M \sqrt{M / \lambda_{1n}} + \mu_{1n} \Omega_{12} + M \simeq \lambda_{1n}^{-1/2}$, we have $\norm{g}_{\infty} \le B$.

    Define the modulus of continuity of $\cG$ by
    \[ \omega_n (\cG, \epsilon) := \bbE \prth{\sup_{g \in \cG, \bbE g^2 \le \epsilon} \abs{\bbE g - \bbP_n g}}, \quad \epsilon > 0, \]
    where the supremum is measurable by the separability assumption on $\cG$.
    Define the function class 
    \begin{equation} \label{equ.def.classE}
        \cE := \brce{\bbE g - g: g \in \cG}, 
    \end{equation}
    and then we have $\omega_n (\cG, 4 B^2) \ge \bbE \sup_{h \in \cE} \bbP_n h$
    since $\abs{\bbE g - g} \le 2B$.
    By Theorem~\ref{thm.Talagrand} we obtain
    \begin{equation} \label{equ.Talagrand.modulus.cont}
        \bbP \prth{\sup_{h \in \cE} \bbP_n h > 3 \omega_n (\cG, 4 B^2) + \sqrt{\frac{2 \tau (4 B)^2}{n}} + \frac{2 B \tau}{n}} \le e^{- \tau}.
    \end{equation}
    Let $\bepsilon = \{ \epsilon_i \}_{i=1}^n$ be a sequence of i.i.d. Rademacher variable.
    Then the local Rademacher average of $\cG$ is defined by
    \[ \Rad(\cG, n, \epsilon) := \bbE_\bz \bbE_{\bepsilon} \sup_{g \in \cG, \bbE g^2 \le \epsilon} \abs{\frac{1}{n} \sum_{i=1}^n \epsilon_i g(\bz_i)}. \]
    It has been shown that
    \[ \omega_n (\cG, \epsilon) \le 2 \Rad(\cG, n, \epsilon), \quad \epsilon > 0 \]
    by symmetrization \citep{vaart1996weak}.
    Since 
    \[ \Rad(\cG, n, \epsilon) = B \Rad(B^{-1} \cG, n, B^{-2} \epsilon) \]
    for any $a > 0$
    by equation (37) of \cite{steinwart2007fast},
    we only need to obtain a bound for $\Rad(B^{-1} \cG, n, B^{-2} \epsilon)$.
    To this end, we will use Proposition 5.5 of \cite{steinwart2007fast} to bound the local Rademacher average, quoted as follows.

    \begin{thm}
        Let $\cF$ be a class of measurable functions from $\bZ$ to $[-1, 1]$ which is separable with respect to $\norm{\cdot}_{\infty}$.
    Assume there are constants $a > 0$ and $0 < p < 2$ with 
    \[ \sup_{\bbP_n} \log N (\epsilon, \cF, L_2 (\bbP_n)) \le a \epsilon^{-p} \]
    for all $\epsilon > 0$.
    Then there exists a constant $c_p > 0$ depending only on $p$ such that for all $n \ge 1$ and all $\epsilon > 0$ we have
    \[ \Rad(\cF, n, \epsilon) \le c_p \max \brce{\epsilon^{\frac{1}{2} - \frac{p}{4}} \prth{\frac{a}{n}}^{\frac{1}{2}}, \prth{\frac{a}{n}}^{\frac{2}{2 + p}}}. \]
    \end{thm}

    Now we need to find some constants $a > 0$ and $0 < p < 2$ such that
    \[ \sup_{\bbP_n} \log N (\epsilon, B^{-1} \cG, L_2 (\bbP_n)) \le a \epsilon^{-p} \]
    for some $a \ge 1, 0 < p < 2$ and for all $\epsilon > 0$.
    To this end, note that
    \begin{equation*}
    \begin{split}
        & \log N (\epsilon, B^{-1} \cG, L_2 (\bbP_n)) \\
        = & \log N \prth{B^{-1} \brce{\ell_1 \circ f_1 + \ell_2 \circ f_1 + \lambda_{1n} \normn{f_1}_{\cH}^2: f \in B_{\cH} (\sqrt{M / \lambda_{1n}})}, \epsilon, L_2 (\bbP_n)} \\
        \le & \log N \prth{B^{-1} \brce{\ell_1 \circ f_1 + \ell_2 \circ f_1 : f \in B_{\cH} (\sqrt{M / \lambda_{1n}})}, \epsilon, L_2 (\bbP_n)} \\
        & + \log N \prth{B^{-1} \brce{\lambda_{1n} \normn{f_1}_{\cH}^2: f \in B_{\cH} (\sqrt{M / \lambda_{1n}})}, \epsilon, L_2 (\bbP_n)}
    \end{split}
    \end{equation*}
    by the subadditivity of the entropy.
    For the first term on the right-hand side, 
    for any $f_1, f_1^{\prime} \in B_{\cH} (\sqrt{M / \lambda_{1n}})$,
    let $u := B^{-1} (\ell_1 + \ell_2) \circ f_1$ and $u^{\prime} := B^{-1} (\ell_1 + \ell_2) \circ f_1^{\prime}$.
    Since $\ell_1$ and $\ell_2$ are Lipschitz continuous with respect to $f_1$, 
    \[ \norm{u - u^{\prime}}_{L_2 (\bbP_n)} \le B^{-1} \prth{M + \frac{\mu_{1n} \Omega_{12} }{\kappa_{1n}}} \normn{f - f^{\prime}}_{L_2 (\bbP_n)}. \]
    With $u, u^{\prime} \in B^{-1} \brce{\ell_1 \circ f_1 + \ell_2 \circ f_1 : f \in B_{\cH} (\sqrt{M / \lambda_{1n}})}$,
    \begin{equation*}
    \begin{split}
        & \log N \prth{B^{-1} \brce{\ell_1 \circ f_1 + \ell_2 \circ f_1 : f \in B_{\cH} (\sqrt{M / \lambda_{1n}})}, \epsilon, L_2 (\bbP_n)} \\
        \le & \log N \prth{B_{\cH} (\sqrt{M / \lambda_{1n}}), \frac{B \epsilon}{M + \mu_{1n} \Omega_{12} \kappa_{1n}^{-1}}, L_2 (\bbP_n)} \\
        \le & \log N \prth{B_{\cH}, \gamma_{1n}^{-1} \epsilon, L_2 (\bbP_n)},
    \end{split}
    \end{equation*}
    since 
    \[ \frac{B}{\sqrt{M / \lambda_{1n}} (M + \mu_{1n} \Omega_{12} \kappa_{1n}^{-1})} \simeq \frac{1}{1 + \mu_{1n} \kappa_{1n}^{-1} } = \gamma_{1n}^{-1}. \]
    For the second term on the right-hand side, it follows that
    \begin{equation*}
        \log N \prth{B^{-1} \brce{\lambda_{1n} \normn{f_1}_{\cH}^2: f \in B_{\cH} (\sqrt{M / \lambda_{1n}})}, \epsilon, L_2 (\bbP_n)} \le \log \frac{M}{B \epsilon}
    \end{equation*}
    since $\lambda_{1n} \normn{f_1}_{\cH}^2 \le M$ for all $f \in B_{\cH} (\sqrt{M / \lambda_{1n}})$.
    Therefore, we can conclude that
    \[ \log N (\epsilon, B^{-1} \cG, L_2 (\bbP_n)) \le \log N \prth{B_{\cH}, \gamma_{1n}^{-1} \epsilon, L_2 (\bbP_n)} + \log \frac{M}{B \epsilon}. \]
    Theorem 2.1 of \cite{steinwart2007fast} then yields that
    \[ \sup_{\bbP_n} \log N (\epsilon, B^{-1} \cG, L_2 (\bbP_n)) \lesssim \sigma_{1n}^{(1 - \nu/2) (1 + \delta) d} (\gamma_{1n}^{-1} \epsilon)^{- \nu}, \]
    where $\sigma_{1n} > 0$ is the parameter of the Gaussian kernel associated with $\cH$, and $0 < \nu \le 2, \delta > 0, \epsilon > 0$.
    Therefore, we have $a = \sigma_{1n}^{(1 - \nu/2) (1 + \delta) d} \gamma_{1n}^{\nu}$, $p = \nu$ and
    \[ \Rad(\cG, n, \epsilon) \le c_p \max \brce{B^{\frac{p}{2}} \epsilon^{\frac{1}{2} - \frac{p}{4}} \prth{\frac{a}{n}}^{\frac{1}{2}}, B \prth{\frac{a}{n}}^{\frac{2}{2 + p}}}. \]
    With $\epsilon = 4 B^2$, we can bound the modulus of continuity as
    \begin{equation} \label{equ.modulus.cont}
        \omega_n (\cG, \epsilon) \le 2 \Rad(\cG, n, \epsilon) \lesssim B a^{\frac{1}{2}} n^{-\frac{1}{2}} \simeq \lambda_{1n}^{-\frac{1}{2}} \sigma_{1n}^{(1 - \nu/2) (1 + \delta) d} \gamma_{1n}^{\nu} n^{-\frac{1}{2}}.
    \end{equation}

    The definition of $\widehat{f}_{1n}$ yields that
    \[ \bbP_n \brce{\ell_1 \circ \widehat{f}_{1n} + \ell_2 \circ \widehat{f}_{1n} + \lambda_{1n} \normn{\widehat{f}_{1n}}_{\cH}^2 - \brck{\ell_1 \circ f_1^{\dagger} + \ell_2 \circ f_1^{\dagger} + \lambda_{1n} \normn{f_1^{\dagger}}_{\cH}^2}} \le 0. \]
    Therefore,
    \begin{equation*}
    \begin{split}
        & \brck{ \cR (\widehat{f}_{1n}) + \lambda_{1n} \normn{\widehat{f}_{1n}}_{\cH}^2 } - \brck{ \cR (f_1^{\dagger}) + \lambda_{1n} \normn{f_1^{\dagger}}_{\cH}^2 } \\
        = & \bbE \brce{\ell_1 \circ \widehat{f}_{1n} + \ell_2 \circ \widehat{f}_{1n} + \lambda_{1n} \normn{\widehat{f}_{1n}}_{\cH}^2 - \brck{\ell_1 \circ f_1^{\dagger} + \ell_2 \circ f_1^{\dagger} + \lambda_{1n} \normn{f_1^{\dagger}}_{\cH}^2}} \\
        \le & (\bbE - \bbP_n) \brce{\ell_1 \circ \widehat{f}_{1n} + \ell_2 \circ \widehat{f}_{1n} + \lambda_{1n} \normn{\widehat{f}_{1n}}_{\cH}^2 - \brck{\ell_1 \circ f_1^{\dagger} + \ell_2 \circ f_1^{\dagger} + \lambda_{1n} \normn{f_1^{\dagger}}_{\cH}^2}} \\
        \le & \sup_{h \in \cE} \bbP_n h,
    \end{split}
    \end{equation*}
    where $\cE$ is defined in (\ref{equ.def.classE}).
    Note that $\norm{h}_{\infty} \le 2B \lesssim \lambda_{1n}^{-\frac{1}{2}}$ and $\bbE h^2 \le 4 B^2 \lesssim \lambda_{1n}^{-1}$ for all $h \in \cE$.
    Plugging (\ref{equ.modulus.cont}) into (\ref{equ.Talagrand.modulus.cont}) and we have
    \begin{equation} \label{equ:estimation.error.result}
    \begin{split}
        & \brck{ \cR (\widehat{f}_{1n}) + \lambda_{1n} \normn{\widehat{f}_{1n}}_{\cH}^2 } - \brck{ \cR (f_1^{\dagger}) + \lambda_{1n} \normn{f_1^{\dagger}}_{\cH}^2 } 
        \le \sup_{h \in \cE} \bbP_n h \\
        \lesssim & \lambda_{1n}^{-\frac{1}{2}} \sigma_{1n}^{(1 - \nu/2) (1 + \delta) d} \gamma_{1n}^{\nu} n^{-\frac{1}{2}}
        + \sqrt{\frac{2 \tau \lambda_{1n}^{-1}}{n}} + \frac{\tau \lambda_{1n}^{-\frac{1}{2}}}{n} \\
        \lesssim & \lambda_{1n}^{-\frac{1}{2}} n^{-\frac{1}{2}} \brck{\sqrt{\tau} + \sigma_{1n}^{(1 - \nu/2) (1 + \delta) d} \gamma_{1n}^{\nu}}
    \end{split}
    \end{equation}
    with probability at least $1 - e^{-\tau}$ for any $\tau \ge 1$.

    Finally, plug the upper bounds (\ref{equ:estimation.error.result}) and (\ref{equ:approx.error.result}) into (\ref{equ:estimation.error}) and (\ref{equ:approx.error}) and we get the results.
\end{proof}

\subsection{Proof of Theorem~\ref{thm:convergence.rate.disagree}} \label{suppsec:sub.proof.disagree}

\begin{proof}
    For $\ell_2$ with the ramp loss, 
    note that $\psi_{\kappa_n} (f_1^* f_2^*) = \bbone \{f_1^* f_2^* < 0\}$ since $f_1^* f_2^*$ takes values only in $\{-1, 1\}$ when $\kappa_n \le 1$.
    Besides, $\psi_{\kappa_n} (f_1 f_2^*) \ge \bbone \{f_1 f_2^* < 0\}$ for any $f_1$ by the definition of the ramp loss $\psi$.
    Hence, we obtain the relationship between the excess risks under the ramp loss and the 0-1 loss as
    \begin{equation} \label{equ:ramploss.01loss.1}
        \bbE \psi_{\kappa_n} (f_1 f_2^*) - \bbE \psi_{\kappa_n} (f_1^* f_2^*) \ge \bbE \bbone \{f_1 f_2^* < 0\} - \bbE \bbone \{f_1^* f_2^* < 0\}. 
    \end{equation}
    Since $\widetilde{f}_2$ and $f_2^*$ are binary decision functions,
    \begin{equation} \label{equ:ramploss.f2}
        \absn{\psi_{\kappa_n} (f_1 \widetilde{f}_2) - \psi_{\kappa_n} (f_1 f_2^*)} 
        = \bbone \{\widetilde{f}_2 f_2^* < 0\} \absn{\psi_{\kappa_n} (f_1 \widetilde{f}_2) - \psi_{\kappa_n} (f_1 f_2^*)} 
        \le \bbone \{\widetilde{f}_2 f_2^* < 0\} 
        \le \widetilde{\delta}_{2 N_2}(\tau)
    \end{equation}
    with probability at least $1 - e^{\tau}$
    for any $f_1$
    by Assumption~\ref{asp:conv.secondary}.
    The first inequality comes from the fact that one of $\psi_{\kappa_n} (f_1 \widetilde{f}_2)$ and $\psi_{\kappa_n} (f_1 f_2^*)$ must be zero and the other is bounded by one.
    Therefore, we have
    \begin{equation} \label{equ:ramploss.01loss.disagreement}
    \begin{split}
        \bbE \ell_2 \circ f_1 - \bbE \ell_2 \circ f_1^* 
        = & \mu_{1n} \Omega_{12} \brck{\bbE \psi_{\kappa_n} (f_1 \widetilde{f}_2) - \bbE \psi_{\kappa_n} (f_1^* \widetilde{f}_2)} \\
        \ge & \mu_{1n} \Omega_{12} \brck{\bbE \psi_{\kappa_n} (f_1 f_2^*) - \bbE \psi_{\kappa_n} (f_1^* f_2^*) - 2 \widetilde{\delta}_{2 N_2}(\tau)} \\
        \ge & \mu_{1n} \Omega_{12} \brck{\bbE \bbone \{f_1 f_2^* < 0\} - \bbE \bbone \{f_1^* f_2^* < 0\} - 2 \widetilde{\delta}_{2 N_2}(\tau)},
    \end{split}
    \end{equation}
    where the first inequality comes from (\ref{equ:ramploss.f2}) and the second inequality comes from (\ref{equ:ramploss.01loss.1}).

    Since 
    $\bbE \ell_1 \circ \widehat{f}_{1n} - \bbE \ell_1 \circ f_1^* + \bbE \ell_2 \circ \widehat{f}_{1n} - \bbE \ell_2 \circ f_1^*
        \lesssim \delta_{1n}(\tau)$,
    we have
    \begin{equation*}
        \mu_{1n} [\bbE \bbone \{\widehat{f}_{1n} f_2^* < 0\} - \bbE \bbone \{f_1^* f_2^* < 0\}
        - \widetilde{\delta}_{2 N_2}(\tau) ] 
        \lesssim \bbE \ell_2 \circ \widehat{f}_{1n} - \bbE \ell_2 \circ f_1^*
        \lesssim \delta_{1n}(\tau)
    \end{equation*}
    with probability at least $1 - 2 e^{\tau}$,
    and the results in Theorem~\ref{thm:convergence.rate.disagree} follows.
\end{proof}

\subsection{Proof of Theorem~\ref{thm:convergence.rate.value}} \label{suppsec:sub.proof.value}

\begin{proof}
    Define $\widehat{U}_{1n} := \bbE \bbone \{\widehat{f}_{1n} f_1^* < 0\}$ and $\Delta \cV_1 (f_1) := \cV_1 (f_1^*) - \cV_1 (f_1)$ for simplicity.

    To utilize existing results in general classification problems, we can rewrite our loss functions with a change of measure.
    Let $h(\cdot)$ be the probability distribution function of the covariates $\bX$.
    Then the expectation of $\ell_1$ can be written as
    \[ \bbE \ell_1 \circ f_1 = \bbE \brck{\frac{R_1}{\pi (A; \bX)} \phi (A f_1 (\bX))} 
    = \int_{\cX_+} \sum_{a \in \cA} \frac{r_1^{(a)} (\bx)}{\pi(a ; \bx)} \phi (a f_1 (\bx)) \pi(a ; \bx) h(\bx) d \bx, \]
    where $\cX_+ := r_1^{(1)} (\bx) + r_1^{(-1)} (\bx) > 0$.
    Now define $g (\bx) := \sum_{a \in \cA} r_1^{(a)} (\bx), C_{R_1} := \int g(\bx) h(\bx) d \bx$. 
    Let $h^{\prime} (\bx) := g(\bx) h(\bx) / C_{R_1}$ so that $h^{\prime}$ is a new probability distribution function.
    Let 
    \begin{equation*}
    \pi^{\prime} (a ; \bx) :=
    \begin{cases}
        \frac{r_1^{(a)} (\bx)}{g(\bx)}, & \text{if } g (\bx) > 0 \\
        \frac{1}{2}, & \text{otherwise},
    \end{cases}
    \end{equation*}
    so that $\pi^{\prime} \in [0, 1]$ by Assumption~\ref{asp:bounded.reward} can be regarded as a new policy for sampling the treatments.
    Then we obtain
    \begin{equation*}
    \begin{split}
        \bbE \ell_1 \circ f_1 
        =& C_{R_1} \int_{\cX_+} \sum_{a \in \cA} \phi (a f_1 (\bx)) \frac{r_1^{(a)} (\bx)}{g(\bx)} \frac{g(\bx) h(\bx)}{C_{R_1}} d \bx \\
        =& C_{R_1} \int \sum_{a \in \cA} \phi (a f_1 (\bx)) \pi^{\prime} (a ; \bx) h^{\prime} (\bx) d \bx.
    \end{split}
    \end{equation*}
    Denote $\bbE^{\prime}$ as the expectation corresponding to the distributions $h^{\prime}$ and $\pi^{\prime}$, so we get 
    \[ \bbE \ell_1 \circ f_1 = C_{R_1} \bbE^{\prime} \phi(A f_1). \]
    Conversely, for the 0-1 loss, the difference between value functions can be written as 
    \begin{equation*}
    \begin{split}
        \Delta \cV_1 (f_1)
        = & \cV_1 (f_1^*) - \cV_1 (f_1) \\
        = & \bbE \brck{\frac{R_1}{\pi (A; \bX)} \bbone \{A f_1 (\bX) < 0\}} - \bbE \brck{\frac{R_1}{\pi (A; \bX)} \bbone \{A f_1^* (\bX) < 0\}} \\
        = & C_{R_1} \brck{\bbE^{\prime} \bbone \{A f_1 < 0\} - \bbE^{\prime} \bbone \{A f_1^* < 0\}}.
    \end{split}
    \end{equation*}
    By Theorem 3 of \cite{bartlett2006convexity} and Assumption~\ref{asp:Tsybakov.noise},
    \begin{equation} \label{equ:logisloss.01loss}
    \begin{split}
        & \bbE \ell_1 \circ f_1 - \bbE \ell_1 \circ f_1^*
        = C_{R_1} \bbE^{\prime} \phi(A f_1) - C_{R_1} \bbE^{\prime} \phi(A f_1^*) \\
        \ge & C_{R_1} c \brck{\bbE^{\prime} \bbone \{A f_1 < 0\} - \bbE^{\prime} \bbone \{A f_1^* < 0\}}^{\alpha} \rho \prth{\frac{\brck{\bbE^{\prime} \bbone \{A f_1 < 0\} - \bbE^{\prime} \bbone \{A f_1^* < 0\}}^{1 - \alpha}}{2c}} \\
        = & C_{R_1} c \brck{\frac{1}{C_{R_1}} \Delta \cV_1 (f_1)}^{\alpha} \rho \brck{\frac{1}{2c} \prth{\frac{1}{C_{R_1}} \Delta \cV_1 (f_1)}^{1 - \alpha}} \\
        \simeq & \brck{\Delta \cV_1 (f_1)}^{\alpha} \rho \brck{\prth{\Delta \cV_1 (f_1)}^{1 - \alpha}}
    \end{split}
    \end{equation}
    for any $f_1$,
    where $c > 0$ is a constant and $\rho (t) = \frac{1}{2} \brck{(1 + t) \log (1 + t) + (1 - t) \log (1 - t)}$ for the logistic loss $\phi$.

    Finally, combine Lemma~\ref{lem:surrogate.risk} with (\ref{equ:ramploss.01loss.disagreement}), (\ref{equ:logisloss.01loss}) and we get that
    \begin{equation*}
    \begin{split}
        & \brck{\Delta \cV_1 (\widehat{f}_{1n})}^{\alpha} \rho \brck{\prth{\Delta \cV_1 (\widehat{f}_{1n})}^{1 - \alpha}}
        + \mu_{1n} [\bbE \bbone \{\widehat{f}_{1n} f_2^* < 0\} - \bbE \bbone \{f_1^* f_2^* < 0\}
        - \widetilde{\delta}_{2 N_2}(\tau) ] \\
        \lesssim &  \bbE \ell_1 \circ \widehat{f}_{1n} - \bbE \ell_1 \circ f_1^* + \bbE \ell_2 \circ \widehat{f}_{1n} - \bbE \ell_2 \circ f_1^*
        \lesssim \delta_{1n}(\tau)
    \end{split}
    \end{equation*}
    with probability at least $1 - 2 e^{\tau}$, that is,
    \begin{equation} \label{equ:final.bound}
    \begin{split}
        \brck{\Delta \cV_1 (\widehat{f}_{1n})}^{\alpha} \rho \brck{\prth{\Delta \cV_1 (\widehat{f}_{1n})}^{1 - \alpha}}
        + \mu_{1n} [\bbE \bbone \{\widehat{f}_{1n} f_2^* < 0\} - \bbE \bbone \{f_1^* f_2^* < 0\}]
        \lesssim \delta_{1n}(\tau) + \mu_{1n} \widetilde{\delta}_{2 N_2}(\tau).
    \end{split}
    \end{equation}
    By Taylor's expansion, it is easy to see that $\rho(t) \simeq t^2$.
    Since $\bbE \bbone \{f_1^* f_2^* < 0\} \le 1$, we can conclude that
    \begin{equation} \label{equ:delta.value}
        \brck{\Delta \cV_1 (f_1)}^{2 - \alpha}
        \lesssim \delta_{1n}(\tau) 
        + \mu_{1n} \widetilde{\delta}_{2 N_2}(\tau)
        + \mu_{1n}
    \end{equation}
    and thus
    \[ \Delta \cV_1 (f_1) \lesssim (\delta_{1n}(\tau) + \mu_{1n})^{\frac{1}{2 - \alpha}} \]
    with probability at least $1 - 2 e^{\tau}$.
\end{proof}

\subsection{Proof of Corollary~\ref{thm:convergence.rate.accuracy}} \label{suppsec:sub.proof.accuracy}

\begin{proof}
    According to Lemma 5 and (9) of \cite{bartlett2006convexity},
    \[ \bbE \bbone \{f_1 f_1^* < 0\} \le c \brck{\bbE(\bbone \{f_1 f_1^* < 0\} \abs{2 \eta(\bX) - 1})}^{\alpha} \]
    where $c$ is some constant.
    By Assumption~\ref{asp:nonzero.rewardsum}, we have
    \[ 2 \eta(\bX) - 1 := \frac{r_1^{(1)} (\bX) - r_1^{(-1)} (\bX)}{r_1^{(1)} (\bX) + r_1^{(-1)} (\bX)} \le \frac{1}{c_r} \abs{r_1^{(1)} (\bX) - r_1^{(-1)} (\bX)}. \]
    Since
    \begin{equation*}
    \begin{split}
        \Delta \cV_1 (f_1)
        = & \bbE \brck{\frac{R_1}{\pi (A; \bX)} \bbone \{A f_1 (\bX) < 0\}} - \bbE \brck{\frac{R_1}{\pi (A; \bX)} \bbone \{A f_1^* (\bX) < 0\}} \\
        = & \bbE \brce{\bbone \{f_1 f_1^* < 0\} \abs{r_1^{(1)} (\bX) - r_1^{(-1)} (\bX)} },
    \end{split}
    \end{equation*}
    we can bound the disagreement rate by the value difference such that
    \begin{equation} \label{equ:accuracy.value.relation}
        \bbE \bbone \{f_1 f_1^* < 0\}
        \le c \brck{\frac{1}{c_r} \bbE(\bbone \{f_1 f_1^* < 0\} \abs{r_1^{(1)} (\bX) - r_1^{(-1)} (\bX)})}^{\alpha} 
        \lesssim \brck{\Delta \cV_1 (f_1)}^{\alpha},
    \end{equation}
    that is, $\widehat{U}_{1n} \lesssim \brck{\Delta \cV_1 (\widehat{f}_{1n})}^{\alpha}$.
    Then following (\ref{equ:delta.value}) we have
    \[ \widehat{U}_{1n}^{\frac{2 - \alpha}{\alpha}}
    \lesssim \delta_{1n}(\tau)
    + \mu_{1n} \widetilde{\delta}_{2 N_2}(\tau)
    + \mu_{1n}
    \]
    with probability at least $1 - 2 e^{\tau}$ if we take $\rho(t) \simeq t^2$.
\end{proof}

\subsection{Proof of Theorems~\ref{thm:convergence.rate.K} and~\ref{thm:convergence.rate.value.accuracy.K}} \label{suppsec:sub.proof.K}

\begin{proof}
    The proof is similar to that for $K = 2$. We only highlight the main differences here.

    To extend the results of Lemma~\ref{lem:surrogate.risk} to $K \ge 3$,
    we can write the surrogate loss for the fusion penalty as $\ell_2 \circ f_1 (\bZ):= \mu_{1n} \sum_{k=2}^K \Omega_{1k} \psi_{\kappa_{1n}} [f_1 (\bX) \widetilde{f}_k (\bX)]$.
    Then $\ell_2$ is Lipschitz continuous with respect to $f_2$ with Lipschitz constant $\frac{\mu_{1n} \sum_{k=2}^K \Omega_{1k}}{\kappa_{1n}}$.
    Hence, we have
    \[ B := 2 M \sqrt{M / \lambda_{1n}} + \mu_{1n} \sum_{k=2}^K \Omega_{1k} + M \simeq \lambda_{1n}^{-1/2} \]
    and 
    \[ \gamma_{1n}^{-1} := \frac{B}{\sqrt{M / \lambda_{1n}} (M + \mu_{1n} \sum_{k=2}^K \Omega_{1k} \kappa_{1n}^{-1})} \simeq \frac{1}{1 + \mu_{1n} \kappa_{1n}^{-1} }, \]
    which shows that the conclusion in Lemma~\ref{lem:surrogate.risk} still holds.

    Now inequality (\ref{equ:ramploss.01loss.disagreement}) should be written as
    \begin{equation} \label{equ:ramploss.01loss.disagreement.K}
    \begin{split}
        \bbE \ell_2 \circ f_1 - \bbE \ell_2 \circ f_1^* 
        = & \mu_{1n} \sum_{k=2}^K \Omega_{1k} \brck{\bbE \psi_{\kappa_n} (f_1 \widetilde{f}_k) - \bbE \psi_{\kappa_n} (f_1^* \widetilde{f}_k)} \\
        \ge & \mu_{1n} \sum_{k=2}^K \Omega_{1k} \brck{\bbE \bbone \{f_1 f_k^* < 0\} - \bbE \bbone \{f_1^* f_k^* < 0\} - 2 \widetilde{\delta}_{k N_k}(\tau)},
    \end{split}
    \end{equation}
    with probability at least $1 - (K - 1) e^{\tau}$
    for any $f_1$
    by Assumption~\ref{asp:conv.secondary}.
    Therefore, inequality (\ref{equ:final.bound}) is changed to
    \begin{equation} \label{equ:final.bound.K}
    \begin{split}
        \brck{\Delta \cV_1 (\widehat{f}_{1n})}^{\alpha} \rho \brck{\prth{\Delta \cV_1 (\widehat{f}_{1n})}^{1 - \alpha}}
        + \mu_{1n} \sum_{k=2}^K [\bbE \bbone \{\widehat{f}_{1n} f_k^* & < 0\} - \bbE \bbone \{f_1^* f_k^* < 0\}] \\
        & \lesssim \delta_{1n}(\tau) + \mu_{1n} \sum_{k=2}^K \widetilde{\delta}_{k N_k}(\tau)
    \end{split}
    \end{equation}
    with probability at least $1 - K e^{\tau}$, and thus
    \begin{equation} \label{equ:delta.value.K}
        \brck{\Delta \cV_1 (f_1)}^{2 - \alpha}
        \lesssim \delta_{1n}(\tau) 
        + \mu_{1n} \sum_{k=2}^K \widetilde{\delta}_{k N_k}(\tau)
        + \mu_{1n}.
    \end{equation}
    The inequality (\ref{equ:convergence.rate.value.K}) follows from the assumption that $\sum_{k=2}^K \widetilde{\delta}_{k N_k}(\tau) = o(1)$.
    
    Similarly, from (\ref{equ:final.bound.K}) we can conclude that 
    \begin{equation*}
        \mu_{1n} [\bbE \bbone \{\widehat{f}_{1n} f_k^* < 0\} - \bbE \bbone \{f_1^* f_k^* < 0\}]
        \lesssim \delta_{1n}(\tau) + \mu_{1n} \sum_{k=2}^K \widetilde{\delta}_{k N_k}(\tau)
    \end{equation*}
    for any $k = 2, \dots, K$ with probability at least $1 - K e^{\tau}$.

    Combining results in (\ref{equ:accuracy.value.relation}) and (\ref{equ:delta.value.K}), we obtain the bound 
    \[ \widehat{U}_{1n}^{\frac{2 - \alpha}{\alpha}}
        \lesssim \delta_{1n}(\tau) 
        + \mu_{1n} \sum_{k=2}^K \widetilde{\delta}_{k N_k}(\tau)
        + \mu_{1n}
    \]
    for the misclassification rate with probability at least $1 - K e^{\tau}$.
\end{proof}

\bibliography{bibfile}